\documentclass{raa_twocolumn}  
\usepackage{graphicx,times}
\usepackage{natbib}
\usepackage{amssymb,amsmath}
\bibpunct{(}{)}{;}{a}{}{,}

\newcommand\targeta{\textrm{J0254}}
\newcommand\targetb{\textrm{J0121}}

\usepackage[pagebackref=true]{hyperref}

\begin{document}

   \title{A photometric study of two contact binaries: CRTS J025408.1+265957 and CRTS J012111.1+272933}

 \volnopage{ {\bf 20XX} Vol.\ {\bf X} No. {\bf XX}, 000--000}
   \setcounter{page}{1}

   \author{Shuo Ma\inst{1,2}, Jin-Zhong Liu\inst{1,2}, Yu Zhang\inst{1,2}, Qing-Shun Hu\inst{1,2}, Guo-Liang L$\rm{\ddot{u}}$\inst{3}
   }

   \institute{ Xinjiang Astronomical Observatory, Chinese Academy of Sciences, Urumqi, Xinjiang 830011, People's Republic of China; {\it liujinzh@xao.ac.cn}\\ \and University of Chinese Academy of Sciences, Beijing 100049, People's Republic of China\\ \and School of Physical Science and Technology, Xinjiang University, Urumqi, 830064, People’s Republic of China\\
\vs \no
   {\small Received 20XX Month Day; accepted 20XX Month Day}
}

\abstract{
We performed new photometric observations for two contact binaries (i.e., CRTS J025408.1+265957 and CRTS J012111.1+272933), which were observed by the 1.0-m telescope at Xingjiang Astronomical Observatory. From our light curves and several survey data, we derived several sets of photometric solutions. We found that CRTS J025408.1+265957 and CRTS J012111.1+272933 were A- and W-type W UMa, respectively. The results imply that the spot migrates or disappears in two contact binaries, which were identified by chromospheric activity emissions (e.g. H$_\alpha$ emission) from LAMOST spectra. From the O-C curves, the orbital periods of two contact binaries may be increasing, which is interpreted by the mass transfer from the less massive component to the more massive one. With mass transferring, two contact binaries may evolve from the contact configurations to semi-detected ones as predicted by the theory of thermal relaxation oscillation.
}

   \authorrunning{Ma et al. }            
   \titlerunning{The Study of Two Eclipsing Binaries}  
   \maketitle

%
\section{Introduction} \label{sec:intro}

Eclipsing binaries are some of the most attractive research subjects in stellar astrophysics because they play a significant role in stellar evolution and are currently the most powerful tool to measure stellar parameters. Eclipsing binaries with light variations where there is little difference between primary and secondary eclipsing depths are referred to as W UMa systems. W UMa systems have approximately similar temperatures in both of their components due to a common envelope \citep{2013MNRAS.430.2029Y}. Normally there is contact between the two components, along with a synchronous circular orbit \citep{2006A&A...446..785M}.

W UMa systems, whose spectral classes range generally from F to K, include two late-type dwarf stars that usually exhibit stellar magnetic activities, such as starspots \citep{2019AJ....157...73K}, flares \citep{2020ApJ...892...58H}, chromosphere activity \citep{2021AJ....161...67W} and magnetic activity cycles \citep{2020AJ....160...62H}. The orbital periods in W UMa systems are usually short, about 0.4 days. It is noteworthy that their orbital periods are generally variational, which can be attributed to a number of factors, including the mass transfer between the two components \citep{2006AJ....132.2260H}, magnetic braking effects \citep{1992ApJ...385..621A,1998MNRAS.296..893L}, magnetic cycles \citep{2005A&A...441.1087B}, and the third body \citep{2006epbm.book.....E,2018NewA...59....1M}. W UMa systems were divided into A- and W-types by \citet{1970VA.....12..217B} - the hotter component has more mass in A-type systems, while W-type systems are the opposite.

The photometric data of W UMa binaries have recently been released by many surveys, such as the Wide Angle Search for Planets \citep[SuperWASP, ][]{2003ASPC..294..405S,2010A&A...520L..10B}, the Catalina Sky Survey \citep[CRTS][]{2009ApJ...696..870D,2014ApJS..213....9D}, the All-Sky Automated Survey for Supernovae \citep[ASAS-SN,][]{2018MNRAS.477.3145J,2019MNRAS.486.1907J,2020MNRAS.493.4045J,2019MNRAS.487.5932P}, the Zwicky Transient Facility \citep[ZTF, ][]{2019PASP..131a8002B}, and the Transiting Exoplanet Survey Satellite \citep[TESS,][]{2014SPIE.9143E..20R,2015JATIS...1a4003R}. These surveys provide an opportunity for the study of W UMa binaries.

In this paper, we select two W UMa binaries named CRTS J025408.1+265957 (TIC 436926796, 2MASS 02540811+2659578, LPSEB19)(hereinafter as J0254) and CRTS J012111.1+272933 (TIC 16996819, 2MASS 01211118+2729334, LPSEB31)(hereinafter as J0121) from the catalog of spectral eclipsing binaries \citep{2020ApJS..249...31Y} to investigate their relevant properties. $\targeta$ and $\targetb$ were periodic variable stars \citep{2014ApJS..213....9D} and were defined as typical W UMa binaries \citep{2019MNRAS.486.1907J}. $\targeta$, with an orbital period of 0.311886 days from the International Variable Star Index (VSX\footnote{\url{https://www.aavso.org/vsx/index.php?view=detail.top&oid=364624}}) or an orbital period of 0.3118858 days \citep{2019MNRAS.486.1907J}, is 14.85 or 15.19 mag in V-band mean magnitude, with amplitude of 0.71 or 0.82. Similarly, for $\targetb$, its orbital period is 0.322818 days in VSX\footnote{\url{https://www.aavso.org/vsx/index.php?view=detail.top&oid=362110}} or 0.322819 days in ASAS-SN, with V-band mean magnitude of 14.92 mag and amplitude of 0.53 in VSX, or a V-band mean magnitude of 15.19 mag and amplitude of 0.57 in ASAS-SN.

In our work, we study the parameters, orbital period variation and magnetic activities of the two W UMa binary targets. The paper is organized as follows. In Section \ref{sec:orb}, we describe the information from the observations and data. The orbital periods are studied in section \ref{sec:orb}. The chromospheric activity is researched in section \ref{sec:spectra}. The analysis of the light curve is in Section \ref{sec:photometric}. In section \ref{sec:discussions}, we discuss the properties of the two targets. Finally, a summary is shown in Section \ref{sec:summary}.

\section{Observation and data} \label{sec:data}

\subsection{New Observation and Data Reduction} \label{subsec:New Obs}

$\targeta$ and $\targetb$ were observed in 2020 with the Nanshan One-meter Wide-field Telescope \citep[hereafter NOWT]{2020RAA....20..211B} at the Nanshan station of the Xinjiang Astronomical Observatory, which is equipped with a standard Johnson multi-color filter system (e.g., $UBVRI$). A CCD camera, with the pixels of 4096 $\times$ 4136, 78 arcmin $\times$ 78 arcmin true field, and 1.125 arcsec each pixel scale, is mounted on this telescope.

The Johnson-Cousins $BVRI$ filters were used during our observations, with the medium scan rate mode. The observation details, such as the target name, observation date, exposure time, number of images, and mean error of photometric observation, are listed in Table \ref{tab:log}. For $\targeta$, a total of 492 CCD images are obtained, and another target ($\targetb$) with 1428 CCD images. The observation precision in more than 90\% is better than 0.008 mag. We note that the third observation night of $\targetb$ is slightly less accurate than the other nights (see row 5 of the Mean Error column in Table \ref{tab:log}), which is probably due to the weather.
\begin{table*}
	\caption{Photometric observation log of $\targeta$ and $\targetb$.}
	\centering
	\tiny
	\begin{tabular*}{\hsize}{@{}@{\extracolsep{\fill}}lcccc@{}}
		\hline\hline
		Target  & UT Date      & Exposures(B\,V\,R\,I) & Number(B\,V\,R\,I) & Mean Error(B\,V\,R\,I)  \\
		        & (yyyy-mm-dd) & (s)                &                 & (mmag)               \\   \hline             
$\targeta$ & 2020-11-06   & 140 \, 120 \, 100 \, 130    & 50 \, 49 \, 50 \, 49     & 8.5 \, 5.9 \, 5.1 \, 6.6      \\          
& 2020-12-22   & 96 \, 65 \, 50 \, 60        & 73 \, 73 \, 74 \, 74     & 8.2 \, 5.9 \, 5.3 \, 6.5       \\ \cline{2-5}
& 2020-09-16   & 60 \, 40 \, 35 \, 30        & 83 \, 83 \, 84 \, 84     & 5.2 \, 4.6 \, 4.4 \, 6.6       \\                    
$\targetb$ & 2020-09-17   & 60 \, 40 \, 35 \, 30        & 108 \, 109 \, 109 \, 108 & 5.7 \, 5.1 \, 4.8 \, 6.9       \\         
& 2020-09-21   & 60 \, 40 \, 35 \, 30        & 52 \, 53 \, 53 \, 53     & 6.7 \, 5.9 \, 5.3 \, 8.1      \\             
& 2020-09-20   & 60 \, 40 \, 35 \, 30        & 112 \, 112 \, 112 \, 113 & 5.8 \, 5.0 \, 4.5 \, 6.9       \\ \hline     
	\end{tabular*}
	\label{tab:log}
\end{table*}
The observed CCD images are reduced by the standard aperture photometry package of the Image Reduction and Analysis Facility (IRAF\footnote{\url{http://iraf.noao.edu/}}) in the standard manner. The process includes image trimming, bias subtraction, flat correction, and aperture photometry. The differential photometry method is adopted in our work. The basic information about the variable stars, the comparison stars, and the check stars are compiled in Table \ref{tab:vcch}, and the partial photometric data are displayed in Table \ref{tab:hjddata}.

\begin{table*}	
	\begin{center}
	\scriptsize
	\caption{The coordinates, JHK magnitude of the target stars, comparison stars, and check stars.}
	\begin{tabular*}{\hsize}{@{}@{\extracolsep{\fill}}llccccc@{}}
		\hline\hline
		Targets        & Name                   & $\alpha_{2000}$ & $\beta_{2000}$ & $Mag\_J $&$ Mag\_H$ & $Mag\_K$ \\ \hline
		Variable star  & $\targeta$                & 02 54 08.11                 & +26 59 57.9                & 13.404 & 13.013 & 12.887 \\
		The comparison & 2MASS 02542302+2659315 & 02 54 22.97                 & +26 59 31.7                & 12.306 & 11.989 & 11.883 \\
		The Check      & 2MASS 02541421+2705339 & 02 54 14.16                 & +27 05 34.5                & 13.633 & 13.151 & 13.016 \\ \hline
		Variable star  & $\targetb$                & 01 21 11.18                 & +27 29 33.5                & 13.667 & 13.34  & 13.272 \\
		The comparison & 2MASS 01211247+2731080 & 01 21 12.40                 & +27 31 08.1                & 13.343 & 12.982 & 12.934 \\
		The Check      & 2MASS 01211812+2730019 & 01 21 18.09                 & +27 30 02.1                & 12.766 & 12.188 & 12.047 \\ \hline
	\end{tabular*}
\label{tab:vcch}
	\end{center}
	\tablecomments{0.96\textwidth}{The coordinates and magnitudes of the variable star and the names, coordinates, and magnitudes of the comparison and check stars are determined from Two Micron All Sky Survey \citep[2MASS;][]{2003yCat.2246....0C}.}
\end{table*}

\begin{table*}
	\bc
	\tiny
	\caption{Observational data of the two eclipsing binaries obtained in 2020.}
	\begin{tabular*}{\hsize}{@{}@{\extracolsep{\fill}}lccccccccc@{}}
		\hline\hline
		Name    &Date       & HJD\_$B$   & $\Delta$Mag\_$B$ & HJD\_$V$   & $\Delta$Mag\_$V$ & HJD\_$R$  & $\Delta$Mag\_$R$& HJD\_$I$ & $\Delta$Mag\_$I$    \\
		&           & (2459000+) &                 &(2459000+)       &                 &(2459000+)             &   & (2459000+)   &                \\ \hline
		$\targeta$  & 2020-11-06 & 160.05716497 & 1.967 & 160.05893583 & 1.950 & 160.06250069 & 1.953 & 160.06075295 & 1.881 \\
		&            & ... & ... & ... & ... &  ... & ... &  ... &  ...\\
		&            & 160.45461035 & 1.342 & 160.45674003 & 1.293 & 160.46029327 & 1.258 & 160.45855715 & 1.217 \\ \cline{2-10}
		& 2020-12-22 & 205.99502343 & 1.293 & 205.99463507 & 1.263 & 205.99708356 & 1.243 & 205.99730709 & 1.191 \\
		&            & ... & ... & ... & ... &  ... & ... &  ... &  ...\\
		&            & 206.32797906 & 1.374 & 206.33021230 & 1.352 & 206.33109725 & 1.324 & 206.33215840 & 1.280 \\ \hline
		$\targetb$  & 2020-09-16 & 109.18889038 & 0.180 & 109.19001316 & 0.213 & 109.19086388 & 0.231 & 109.19177254 & 0.238 \\
		&            & ... & ... & ... & ... &  ... & ... &  ... &  ...\\
		&            & 109.45586049 & 0.393 & 109.45667236 & 0.432 & 109.45733632 & 0.393 & 109.45808291 & 0.404 \\ \cline{2-10}
		& 2020-09-17 & 110.11540897 & 0.325 & 110.11592983 & 0.347 & 110.11676898 & 0.368 & 110.12093008 & 0.316 \\
		&            & ... & ... & ... & ... &  ... & ... &  ... &  ...\\
		&            & 110.45853000 & 0.193 & 110.46237136 & 0.227 & 110.46303688 & 0.235 & 110.46378347 & 0.253 \\ \cline{2-10}
		& 2020-09-19 & 112.22555024 & 0.206 & 112.22637205 & 0.246 & 112.22702601 & 0.251 & 112.22777259 & 0.262 \\
		&            & ... & ... & ... & ... &  ... & ... &  ... &  ...\\
		&            & 112.40663539 & 0.170 & 112.40744563 & 0.213 & 112.40811115 & 0.218 & 112.40885768 & 0.223 \\ \cline{2-10}
		& 2020-09-21 & 114.10625343 & 0.739 & 114.11026978 & 0.717 & 114.10790283 & 0.738 & 114.10864935 & 0.725 \\
		&            & ... & ... & ... & ... &  ... & ... &  ... &  ...\\
		&            & 114.45770486 & 0.423 & 114.46191803 & 0.391 & 114.46258361 & 0.378 & 114.46331851 & 0.369  \\ \hline
	\end{tabular*}
	\label{tab:hjddata}
	\ec
	\tablecomments{0.96\textwidth}{This table is available in its entirety in machine-readable form.}
\end{table*}

\subsection{Survey Data} \label{subsec:survey}

The photometric data of $\targeta$ and $\targetb$ are also obtained from the SuperWASP, CRTS, ASAS-SN, and ZTF survey databases. These data are characterized as follows: (a) unfiltered observation and only V-band data available in CRTS; (b) only V-band data in ASAS-SN; (c) ZTF including $g$, $r$ bands for our targets; (d) the data in SuperWASP obtained by multiple times scan in same night. There are big magnitude errors in the SuperWASP data for the two targets, which may be because the two targets are outside the suitable magnitude range of the telescope. Among these data, only the SuperWASP data needs to be processed. Therefore, the raw data of SuperWASP were processed in the following steps: firstly, removing the data with magnitude errors greater than 1 mag, then the remaining data with magnitude errors outside three standard deviations being excluded.

\begin{figure*}
	\centering	
	\includegraphics[width=150mm]{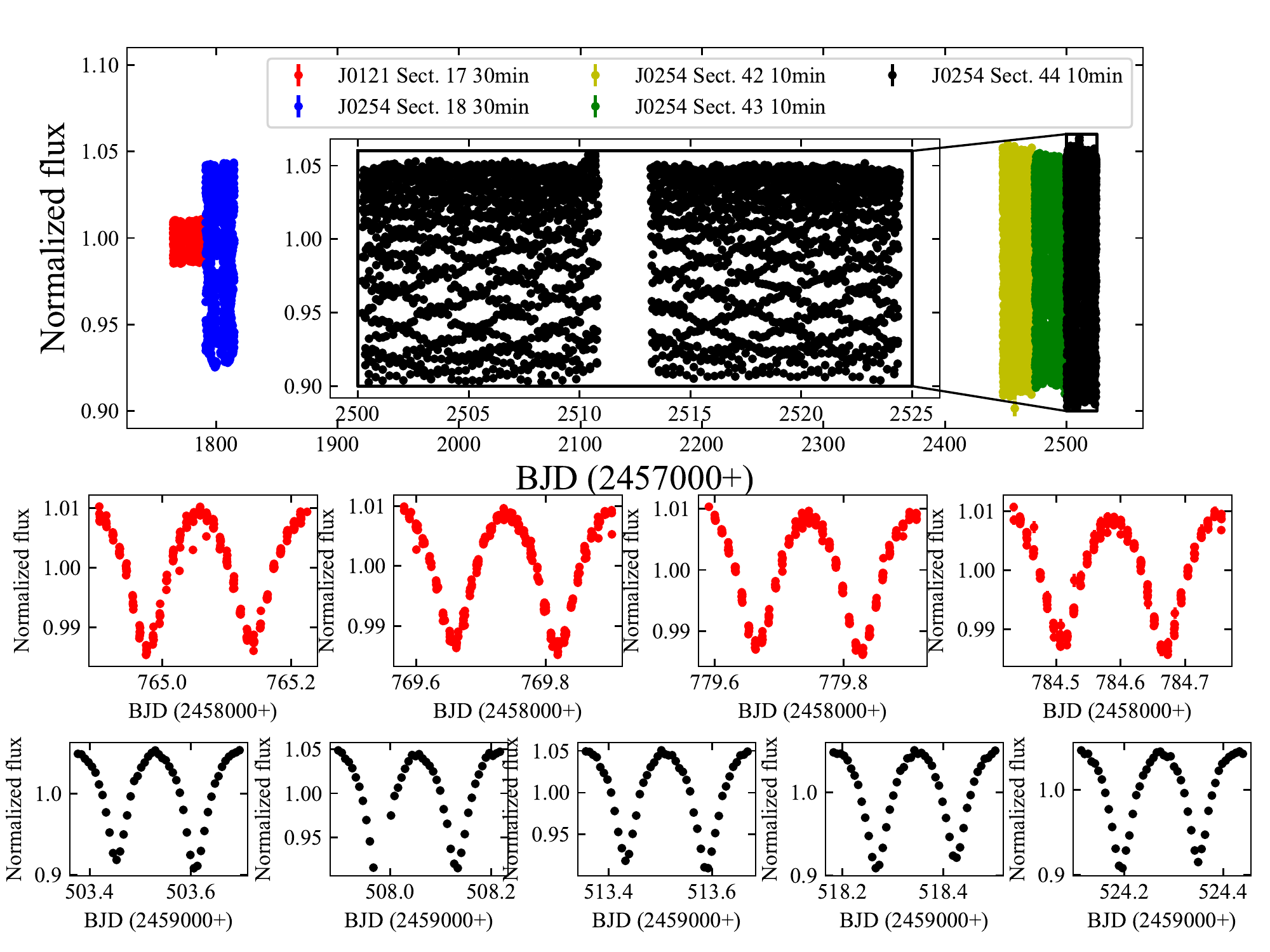}
	\caption{The upper panel shows the TESS data of $\targeta$ (Sectors 18, 42, 43, 44) and $\targetb$ (Sector 17) and the partial enlargement of the data from sector 44. Then, the $\targetb$ data of 30 minutes cadence are divided into four parts, and the data of each part are transformed into one period and displayed in the middle panel. The data of $\targeta$ from sector 44 are divided into five parts and extracted one period of data from each part and shown in the bottom panel.}
	\label{fig:TESS}
\end{figure*}

In addition, the photometric data of the two objects are also found in TESS database. $\targeta$ was observed in Sectors 18, 42, 43, and 44, while $\targetb$ in Sector 17. The principal data products collected by the TESS mission exist in three forms (the Full Frame Images (FFIs), the Target Pixel Files (TPFs), and the Light Curve Files (LCFs)). Only FFIs, with a cadence of 30 minutes (Sectors 17, 18) and 10 minutes (Sectors 42, 43, 44), of both targets were available. We adopted the python package Lightkurve\footnote{\url{http://docs.lightkurve.org/}} to reduce them and obtain the light curves for the two targets by using the PLDCorrector\footnote{\url{https://docs.lightkurve.org/tutorials/2-creating-light-curves/2-3-k2-pldcorrector.html}}. The TESS data are shown in the upper panel of Fig. \ref{fig:TESS}.

In this paper, we aim to seek the times of minima (eclipse timings) based on the data from TESS, which was used for $(O-C)$ analysis. The different methods were used to deal with the data with different cadence. The data with the 30 minutes cadence were clipped and stacked with a similar method adopted by \citet{2021AJ....162...13L} to obtain more times of minima.

For each observation sector, the data of 30 minutes cadence are divided into four parts among which each part was converted into one period through the equation $BJD = BJD_0 + P \times E$, with $BJD$ representing the observing time, $BJD_0$ denoting the reference time, $E$ referring to the cycle, and the orbital period indicated as $P$. In this equation, the orbital periods of $\targeta$ and $\targetb$ we adopted were 0.311886 and 0.322818 days in VSX, respectively. By the method, the corresponding four diagrams of $\targetb$ are obtained, as shown in the middle panel of Fig. \ref{fig:TESS}. We can also obtain times of minima on the data of 10 minutes cadence, which based on the manner, that is, the data for each sector is divided into five parts from which the light curve in one period was extracted each part. For example, we displayed the five light curves of $\targeta$ from sector 44 in the bottom panel of Fig. \ref{fig:TESS}.

Except photometric data, the spectral data of the two targets can be found in LAMOST Data Release 7\footnote{\url{http://www.lamost.org/dr7/}}, with five low-resolution spectra for $\targeta$ and two low-resolution spectra for $\targetb$. Their spectral parameters are tabulated in Table \ref{tab:lamostlog}, which including the object, observational date, Heliocentric Julian date, phase, spectral type, as well as effective temperature, surface gravity, radial velocity, and their corresponding errors. The phase is calculated based on the linear ephemeris (see more details in Section \ref{sec:orb}).

\begin{table*}
	\caption{LAMOST spectral information of $\targeta$ and $\targetb$.}
	\centering
	\tiny
	\begin{tabular*}{\hsize}{@{}@{\extracolsep{\fill}}lcccclccr@{}}
		\hline\hline
		Target  & UT Date     & HJD            &  phase           & Subclass & \multicolumn{1}{c}{$T_{eff}$}   & log $g$     &[Fe/H]    & Radial velocity    \\
		        & (yyyy-mm-dd)& (2450000+)            &                  &          & \multicolumn{1}{c}{(K)}    &             &          &(km$\cdot$s$^{-1}$) \\ \hline
		        & 2011-12-03  & 5899.15799  &  0.05738         & G3       &                 &                   &          &                    \\
		        & 2011-12-11  & 5907.08052  &  0.45941         & F9       &                      &                   &          &                    \\
		$\targeta$ & 2012-01-12  & 5939.01222  &  0.84208         & G3       & 5594 $\pm$ 126 & 4.24 $\pm$ 0.21 & -0.20 $\pm$ 0.12 & -3.2  $\pm$ 27.0 \\
		        & 2013-01-16  & 6308.96913  &  0.03540 & G4   & 5474 $\pm$ 172 & 4.11 $\pm$ 0.27 & -0.22 $\pm$ 0.16 & 8.6   $\pm$ 12.9 \\
		        & 2013-01-17  & 6309.97329  &  0.25504         & G7       & 5565 $\pm$ 146 & 4.18 $\pm$ 0.24 & -0.07 $\pm$ 0.14 & -13.0 $\pm$ 26.2 \\ \cline{2-9}
		$\targetb$ & 2014-11-11  & 9278.76899  &  0.58364         & G3       & 5662 $\pm$ 43  & 4.09 $\pm$ 0.07 & -0.18 $\pm$ 0.04 & -37.9  $\pm$ 16.3 \\
		        & 2014-12-17  & 7008.98636  &  0.42706         & G3       & 5656 $\pm$ 27  & 4.04 $\pm$ 0.05 & -0.26 $\pm$ 0.03 & -37.4  $\pm$ 10.0 \\ \hline
	\end{tabular*}
	\label{tab:lamostlog}
\end{table*}

\section{Orbital Period Study} \label{sec:orb}

In this paper, the higher temperature components of the binary systems were regarded as the primary components. With our observations, the eclipse timings were obtained through the K-W method given by \citet{1956BAN....12..327K} from the $BVRI$-bands light curves. The new times of minima by us and their mean values are shown in Table \ref{tab:our min}.

\begin{table*}
	\tiny
	\caption{Newly obtained eclipse timings of $\targeta$ and $\targetb$ in the $BVRI$ Bands.}
	\begin{tabular*}{\hsize}{@{}@{\extracolsep{\fill}}lllllll@{}}
		\hline\hline
		& HJD(B)                & HJD(V)                & HJD(R)              & HJD(I)              & HJD(average)           & Min.      \\
		& 2459000+              & 2459000+              & 2459000+            & 2459000+            & 2459000+               &           \\ \hline
		$\targeta$ & 160.06470$\pm$0.00016 & 160.06520$\pm$0.00047 &                     &                         & 160.06495$\pm$0.00025 & s \\
		& 160.22085$\pm$0.00092 & 160.22112$\pm$0.00115 & 160.22098$\pm$0.00017 & 160.22072$\pm$0.00027 & 160.22092$\pm$0.00038 & p   \\
		& 160.37665$\pm$0.00048 & 160.37708$\pm$0.00046 & 160.37720$\pm$0.00077 & 160.37685$\pm$0.00055 & 160.37695$\pm$0.00029 & s \\
		& 206.06571$\pm$0.00014 & 206.06514$\pm$0.00030 & 206.06573$\pm$0.00011 & 206.06516$\pm$0.00028 & 206.06544$\pm$0.00011 & p   \\
		& 206.22074$\pm$0.00027 & 206.22080$\pm$0.00006 & 206.22147$\pm$0.00018 & 206.22104$\pm$0.00024 & 206.22101$\pm$0.00010 & s \\ \hline
		$\targetb$ & 109.26328$\pm$0.00016 & 109.26350$\pm$0.00031 & 109.26392$\pm$0.00032 & 109.26362$\pm$0.00074 & 109.26358$\pm$0.00022 & s \\
		& 109.42392$\pm$0.00020 & 109.42385$\pm$0.00022 & 109.42398$\pm$0.00020 & 109.42387$\pm$0.00031 & 109.42391$\pm$0.00012 & p  \\
		& 110.23138$\pm$0.00015 & 110.23131$\pm$0.00008 & 110.23184$\pm$0.00048 & 110.23092$\pm$0.00017 & 110.23136$\pm$0.00013 & s \\
		& 110.39235$\pm$0.00016 & 110.39241$\pm$0.00013 & 110.39241$\pm$0.00015 & 110.39220$\pm$0.00024 & 110.39234$\pm$0.00009 & p   \\
		& 112.32907$\pm$0.00027 & 112.32899$\pm$0.00016 & 112.32922$\pm$0.00014 & 112.32892$\pm$0.00017 & 112.32905$\pm$0.00010 & p   \\
		& 114.26625$\pm$0.00032 & 114.26654$\pm$0.00045 & 114.26641$\pm$0.00038 & 114.26616$\pm$0.00046 & 114.26634$\pm$0.00020 & p  \\
		& 114.42869$\pm$0.00019 & 114.42799$\pm$0.00023 & 114.42789$\pm$0.00012 & 114.42781$\pm$0.00027 & 114.42809$\pm$0.00010 & s \\ \hline
	\end{tabular*}
	\label{tab:our min}
\end{table*}

In order to study the orbital period variations, we searched for as many eclipse timings of $\targeta$ and $\targetb$ as we could. Thirty-six eclipse timings of $\targeta$ and thirty-two eclipse timings of $\targetb$ were obtained from the SuperWASP database by the K-W method in Nelson's program\footnote{\url{https://www.variablestarssouth.org/software-by-bob-nelson/}}. The eclipse timings of $\targeta$ and $\targetb$ can also found from \citet{2020ApJS..249...31Y}, who pointed out that these times are primary eclipse timings. In the subsequent analysis, the times of minima were transformed from MJD to HJD through the website\footnote{\url{http://www.physics.sfasu.edu/astro/javascript/hjd.html}}. In addition, the times of minima of $\targeta$ and $\targetb$ were also provided by the ASAS-SN. There are not corresponding errors in the eclipsing times obtained from \citet{2020ApJS..249...31Y} and ASAS-SN. Therefore, we set arbitrarily their errors to 0.002 in the following analysis. By the K-W method, we also obtain the eclipse timings based on the light curves in one period from TESS data above. These eclipse timings are transformed from BJD to HJD through the website\footnote{\url{https://astroutils.astronomy.osu.edu/time/bjd2utc.html},\citep{2010PASP..122..935E}}\footnote{\url{http://www.physics.sfasu.edu/astro/javascript/hjd.html}}. In brief, we acquired all times of minima between 2004 and 2021, which were displayed in Table \ref{tab:all_min}.

\begin{table*}[htpb]
	\bc
	\scriptsize
	\caption{Eclipse Timings of $\targeta$ and $\targetb$.}
	\begin{tabular*}{\hsize}{@{}@{\extracolsep{\fill}}lccrrrrc@{}}
		\hline\hline
		\multicolumn{1}{c}{Star} & \multicolumn{1}{c}{HJD(2400000+)} & Error   & \multicolumn{1}{c}{E} & \multicolumn{1}{c}{$(O-C)_1$} &  \multicolumn{1}{c}{$(O-C)_2$} & \multicolumn{1}{c}{Residual} & Reference \\ \hline
		$\targeta$ & 53242.65066 &  0.00044 &  -18973.5 &  -0.00503 &   0.00067  & -0.00287  &  (1)  \\
		& 53246.70321 &  0.00083 &  -18960.5 &  -0.00700 &  -0.00130  & -0.00483  &  (1)  \\
		& 53993.67529 &  0.00136 &  -16565.5 &  -0.00141 &   0.00351  &  0.00210  &  (1)  \\
		& 53994.60800 &  0.00134 &  -16562.5 &  -0.00436 &   0.00056  & -0.00085  &  (1)  \\
		& 53995.70804 &  0.00187 &  -16559.0 &   0.00408 &   0.00900  &  0.00759  &  (1)  \\
		& 53997.57440 &  0.00082 &  -16553.0 &  -0.00087 &   0.00404  &  0.00264  &  (1)  \\
		& 54003.65070 &  0.00105 &  -16533.5 &  -0.00635 &  -0.00144  & -0.00283  &  (1)  \\ \hline
	\end{tabular*}
	\label{tab:all_min}
	\ec
	\tablecomments{0.96\textwidth}{(1)This paper (SuperWASP); (2)\citet{2020ApJS..249...31Y}; (3)ASAS-SN; (4)This paper (TESS); (5)This paper (NOWT).\\This table is available in its entirety in machine-readable form.}

\end{table*}

We fit $(O-C)$ diagrams with the OCFit package \citep{2019OEJV..197...71G} in which the robust regression method was employed to achieve linear ephemeris fitting. The formula to indicate the linear ephemeris is as follows:
\begin{equation}
	HJD = HJD_0 + P\times E. \\
\label{eq:linear}
\end{equation}

\begin{figure*}[htpb]
	\centering
        \includegraphics[width=0.48\textwidth]{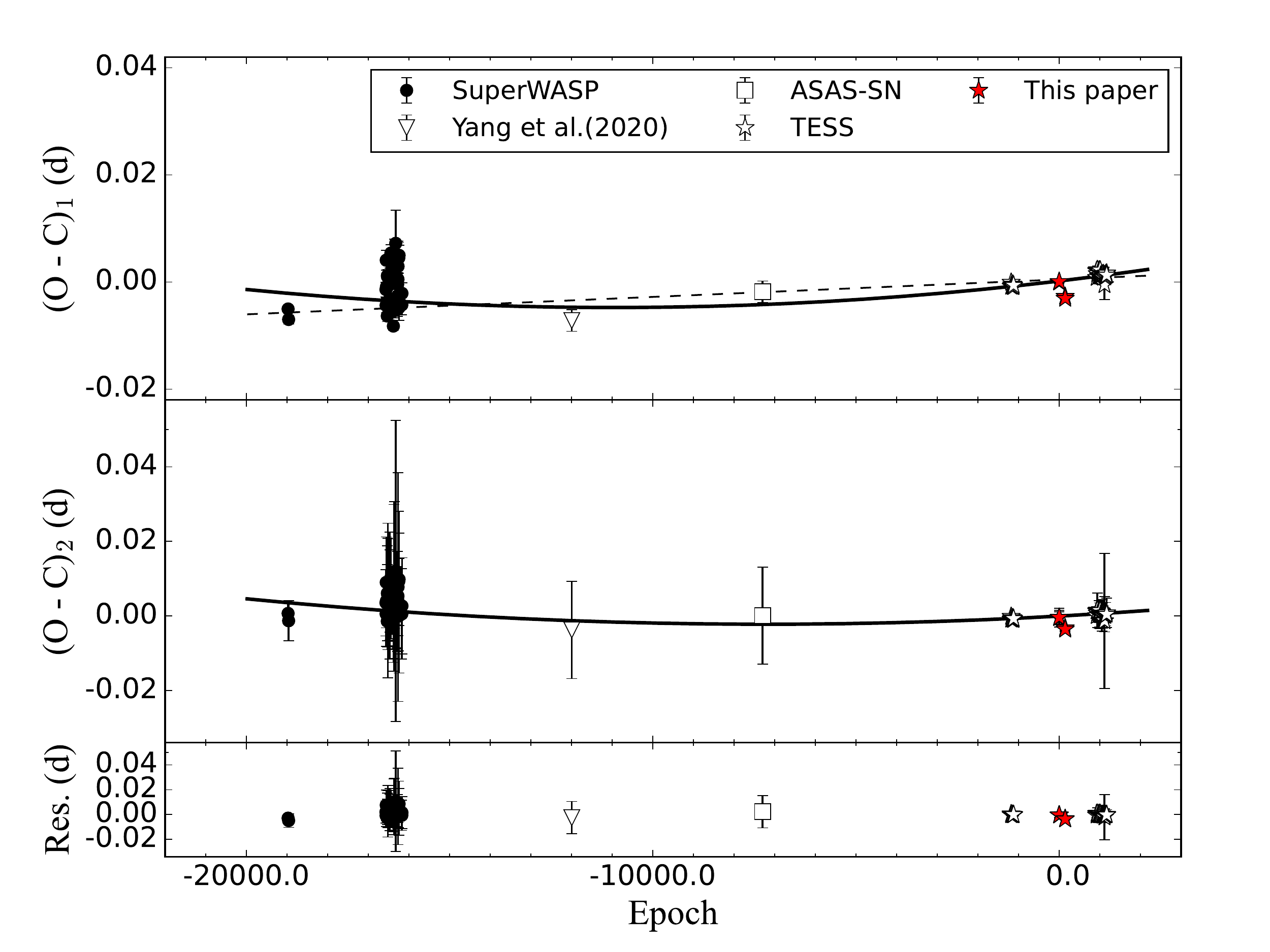}
        \includegraphics[width=0.48\textwidth]{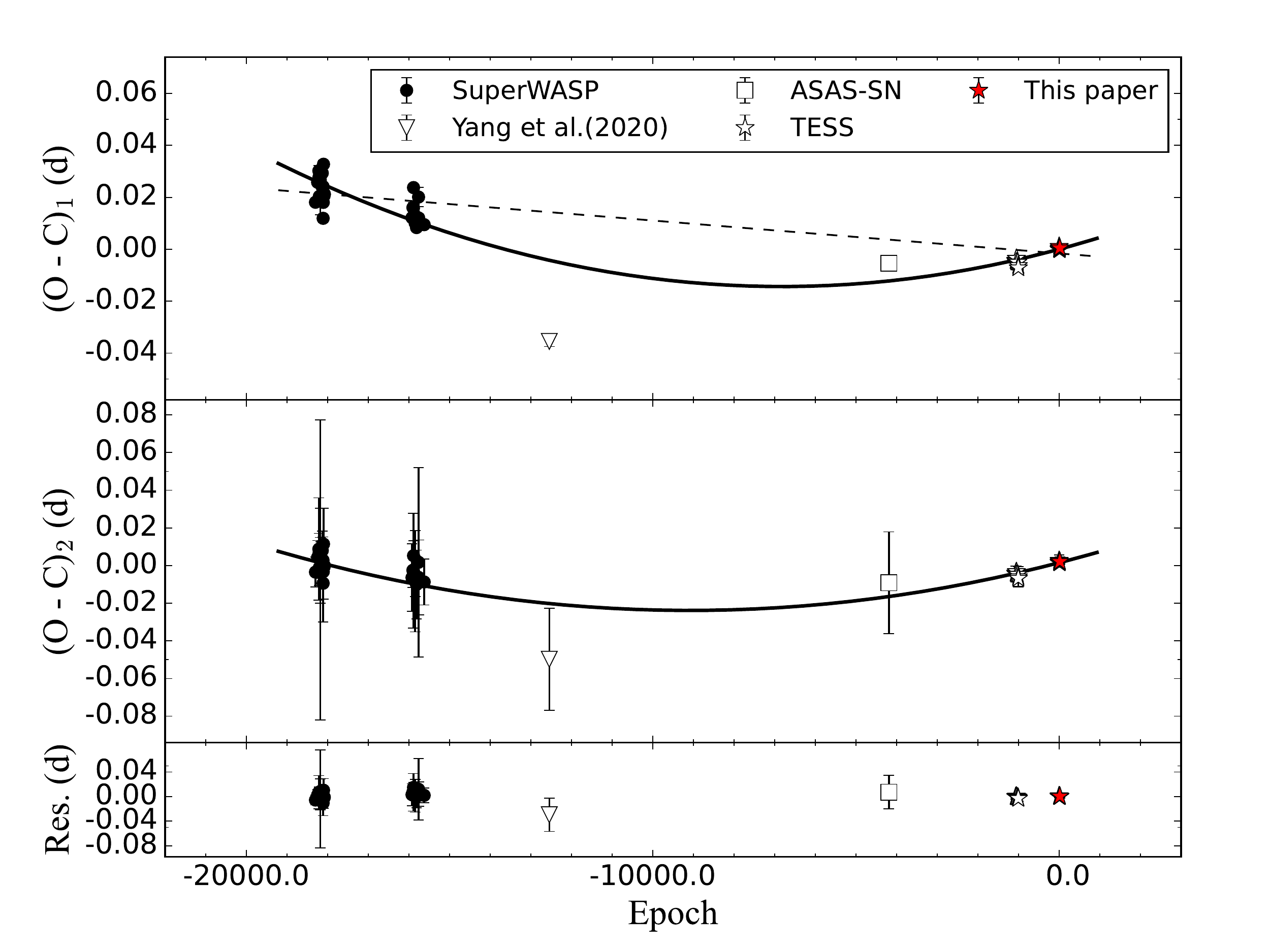}
	\caption{$(O - C)$ diagrams for $\targeta$ (left) and $\targetb$ (right). The upper panel shows the $(O - C)_1$ data based on the initial information, with dotted lines represent linear fitting and solid lines indicate quadratic fitting. In the middle panel, $(O - C)_2$ is plotted using linear ephemeris to remove linear fitting $(O - C)_1$, the solid line shows a quadratic fitting for $(O - C)_2$, and the lower panel shows the quadratic fitting residuals for $(O - C)_2$.}
	\label{fig:ocfit}
\end{figure*}

\subsection{$\targeta$}

For $\targeta$, a total of eighty-one times of minima are determined, which are listed in Table \ref{tab:all_min}. Based on the period of 0.3118858 days from the ASAS-SN database and all times of minima, the first linear ephemeris can be expressed as:
\begin{eqnarray}
	\textrm{Min.I} &=&2459160.22140\,(9) \nonumber \\
	& &+ 0.31188613\,(3) \times E,
\end{eqnarray}

The $(O - C)_2$ values based on this ephemeris plotted in the left panel of Fig. \ref{fig:ocfit} suggests a continuously increasing period. A quadratic fit to these times of minima yields following ephemeris:

\begin{eqnarray}
	\textrm{Min.I} &=&2459160.22135\,(5) \nonumber \\
	& &+ 0.31188673\,(5)\times E  \nonumber \\
	& &+4.2\,(3)\times 10^{-11} \times E^2,
\label{eq:oc1}
\end{eqnarray}
a continuous period increase ratio of $dP/dt = 9.8\,(7)\times10^{-8}$ days yr$^{-1}$ is derived from the quadratic term.

\subsection{$\targetb$}

We obtained forty-nine eclipse timings of $\targetb$. With the period of 0.322819 days given by the ASAS-SN and all eclipse timings in Table \ref{tab:all_min}, we derived the following linear ephemeris:
\begin{eqnarray}
	\textrm{Min.I} &=&2459109.42231\,(54) \nonumber \\
	& &+ 0.32281773\,(7) \times E.
\end{eqnarray}
The $(O-C)$ diagram plotted in the right panel of Fig. \ref{fig:ocfit} clearly shows the signature of
a linearly increasing period. The quadratic ephemeris was obtained as follows:
\begin{eqnarray}
	\textrm{Min.I} &=&2459109.42381\,(19) \nonumber \\
	& &+ 0.32282331\,(31)\times E  \nonumber \\
	& &+ 3.1\,(2)\times 10^{-10} \times E^2,
\label{eq:oc2}
\end{eqnarray}
implying a continuous period increase of $dP/dt = 6.9\,(4)\times10^{-7}$ days yr$^{-1}$.

For the two targets, we note that eclipse timings from \cite{2020ApJS..249...31Y} did not correspond to the primary eclipse timings but secondary eclipse timings. The O-C analysis suggested that the orbital period of both targets is increasing, which will be discussed in Section \ref{subsec:transfer and evolution}.

\section{Chromosphere activity analysis} \label{sec:spectra}

Chromospheric emission lines, including Ca II IRT, H$_{\alpha}$, H$_{\beta}$, H$_{\gamma}$, H$_{\delta}$, and Ca II H$\&$K, are usually used to study whether late-type stars have chromospheric activity \citep{1993ApJS...85..315S,2020ApJS..247....9Z}. Our targets $\targeta$ and $\targetb$ have 5 and 2 low-resolution spectra from LAMOST, respectively. The low-resolution spectra of binary is represented as a common characterization of its two components. The chromosphere activity emission lines of the binary are masked by its photosphere stronger absorption lines. So the spectral subtraction technique, whose principle is that stars with similar spectral types have nearly the same level of photospheric flux, is adopted in this paper to remove the influence of the photosphere. The chromospheric flux of active stars can be estimated by subtracting that of inactive stars with similar to the spectral types of the active stars\citep{1995A&AS..114..287M}.

\begin{figure*}[htpb]
	\centering	
        \includegraphics[width=0.48\textwidth]{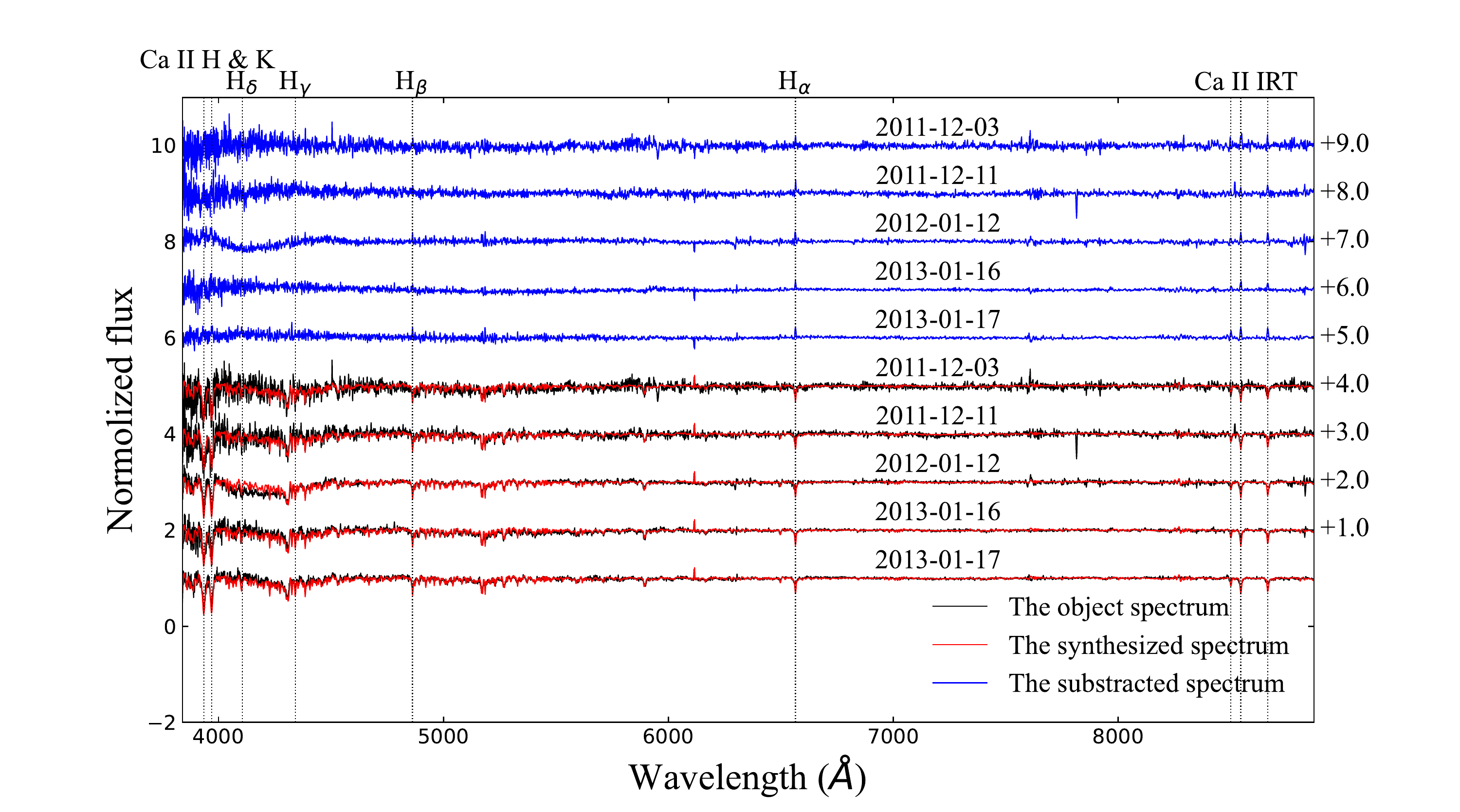}
        \includegraphics[width=0.48\textwidth]{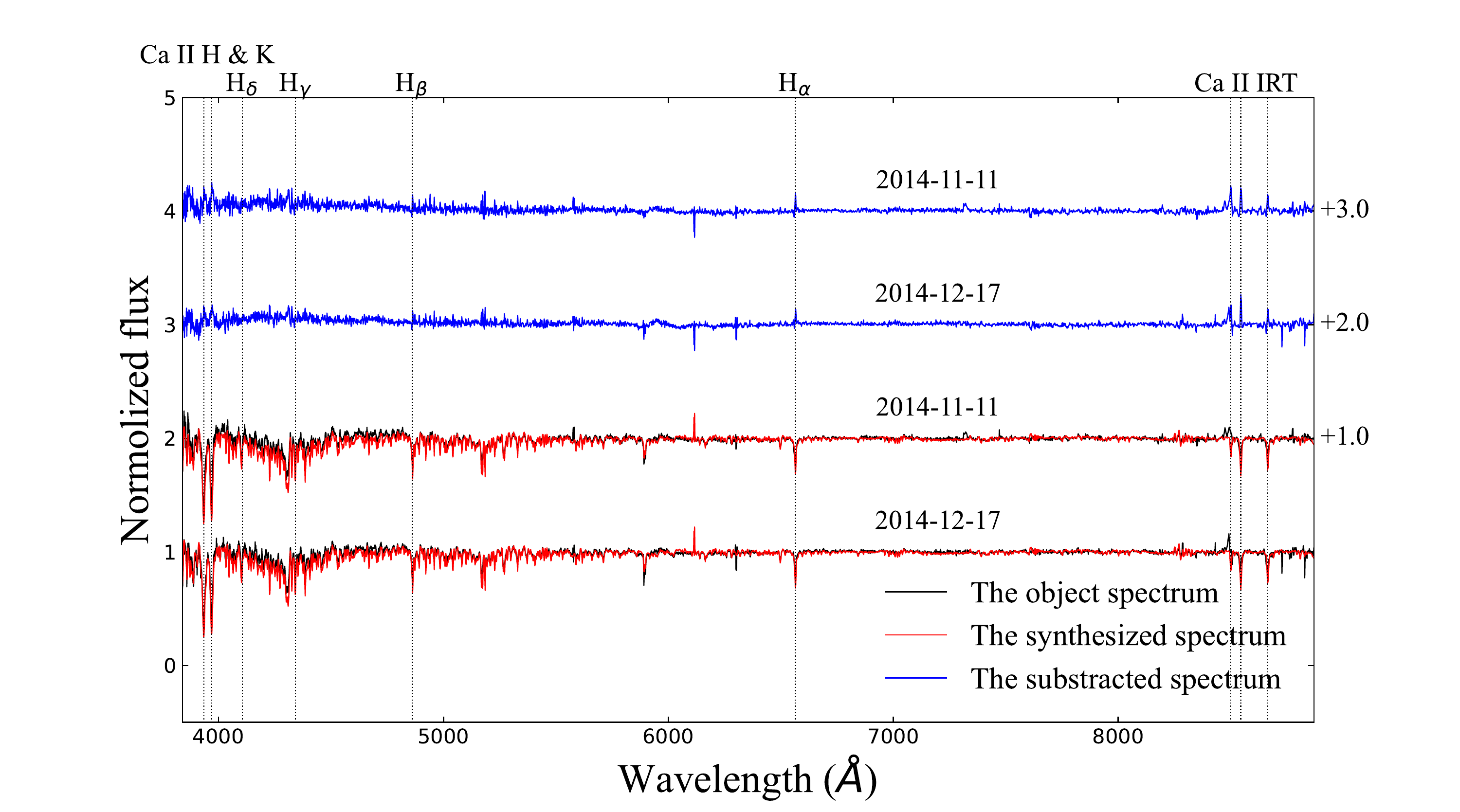}
	\caption{The normalized object, synthesized and subtracted spectra for $\targeta$ (left) and $\targetb$ (right) observed by LAMOST. The excess emission lines are marked with dotted lines.}
	\label{fig:spectra}
\end{figure*}

\citet{2000A&AS..142..275S} provided a list of 750 stars with no chromospheric activity. For these 750 sources, a total of fifty-two spectral lines were obtained from LAMOST DR7. There are thirty-nine spectral lines after removing poor spectral lines. The remaining lines were normalized to cross-match the seven spectral lines of the two targets , then obtained four spectra of HD 224844, HD 87680, and HD 13357. These four spectra were used to make synthetic spectra. The chromosphere activity signals of the two targets are detected by subtracting the synthesized spectra obtained from the spectra of the two targets. The results are shown in Fig. \ref{fig:spectra}.

To evaluate these chromospherically active emission lines, we used the PHEW\footnote{PytHon Equivalent Widths \url{https://zenodo.org/record/47889}} package to calculate the equivalent widths (EWs) of these emission lines. The results obtained after a thousand MC iterations are listed in Table \ref{tab:EWs}. Note that the low-resolution spectra are caused by the coupling of the two components of the two targets. Therefore, the spectral line broadening mechanism makes the equivalent widths corresponding to these emission lines numerically small. The characteristics of the emission lines in Fig. \ref{fig:spectra} correspond essentially to their obtained equivalent widths, implying the presence of chromospheric activity in the both targets.

\begin{table*}[htpb]
	\caption{The EWs of chromospheric activity indicators}
	\centering
	\tiny
	\resizebox{\textwidth}{!}
	{\begin{tabular}{llccccccccc}
		\hline\hline
		Target  & UT Date & Ca II K & Ca II H & H$_{\delta}$ & H$_{\gamma}$ & H$_{\beta}$ & H$_{\alpha}$ & Ca II IRT & Ca II IRT & Ca II IRT  \\
		&& 3934.8{\AA} & 3969.6{\AA} & 4105.9{\AA} & 4341.7{\AA} & 4862.7{\AA} & 6564.6{\AA} & 18500.4{\AA} & 28544.4{\AA} & 38664.5{\AA}  \\  \hline
$\targeta$ & 2011-12-03 & 2.05  $\pm$ 1.15  &0.40  $\pm$ 0.35 & 0.34  $\pm$ 0.23  &      -          & 0.52  $\pm$ 0.24  &0.70  $\pm$ 0.16 &        -      & 2.04  $\pm$ 0.57 & 1.19  $\pm$ 0.17  \\
& 2011-12-11 & 1.29  $\pm$ 0.53  &0.57  $\pm$ 0.34 & 0.45  $\pm$ 0.37  &0.50  $\pm$ 0.25 &       -         &0.77  $\pm$ 0.12 & 0.66  $\pm$ 0.06 & 0.86  $\pm$ 0.24 & 1.44  $\pm$ 0.18  \\
& 2012-01-12 & 0.58  $\pm$ 0.34  &1.21  $\pm$ 0.47 &       -          &0.71  $\pm$ 0.12 & 0.62  $\pm$ 0.09  &1.09  $\pm$ 0.15 & 1.09  $\pm$ 0.26 & 1.56  $\pm$ 0.18 & 1.39  $\pm$ 0.19    \\
& 2013-01-16 & 1.92  $\pm$ 0.66  &0.90  $\pm$ 0.48 & 0.88  $\pm$ 0.39  &0.28  $\pm$ 0.17 & 0.47  $\pm$ 0.17  &0.74  $\pm$ 0.07 & 0.74  $\pm$ 0.18 & 1.26  $\pm$ 0.11 & 0.96  $\pm$ 0.14  \\
& 2013-01-17 & 0.96  $\pm$ 0.36  &1.24  $\pm$ 0.35 & 0.40  $\pm$ 0.06  &0.55  $\pm$ 0.14 & 0.62  $\pm$ 0.14  &1.15  $\pm$ 0.10 & 1.03  $\pm$ 0.25 & 1.87  $\pm$ 0.13 & 1.76  $\pm$ 0.24  \\ \hline
$\targetb$ & 2014-11-11 & 0.51  $\pm$ 0.23  &0.78  $\pm$ 0.33 & 0.30  $\pm$ 0.10  &0.32  $\pm$ 0.07 & 0.45  $\pm$ 0.05  &0.68  $\pm$ 0.11 & 1.31  $\pm$ 0.40 & 1.75  $\pm$ 0.20 & 1.10  $\pm$ 0.21  \\
& 2014-12-17 & 0.49  $\pm$ 0.53  &0.53  $\pm$ 0.14 & 0.11  $\pm$ 0.10  &0.18  $\pm$ 0.06 & 0.29  $\pm$ 0.03  &0.43  $\pm$ 0.07 & 0.74  $\pm$ 0.46 & 1.66  $\pm$ 0.01 & 1.03  $\pm$ 0.27  \\ \hline
	\end{tabular}}
	\label{tab:EWs}
\end{table*}

\section{Photometric analysis} \label{sec:photometric}
\begin{figure*}[htpb]
	\centering
	\includegraphics[width=0.48\textwidth]{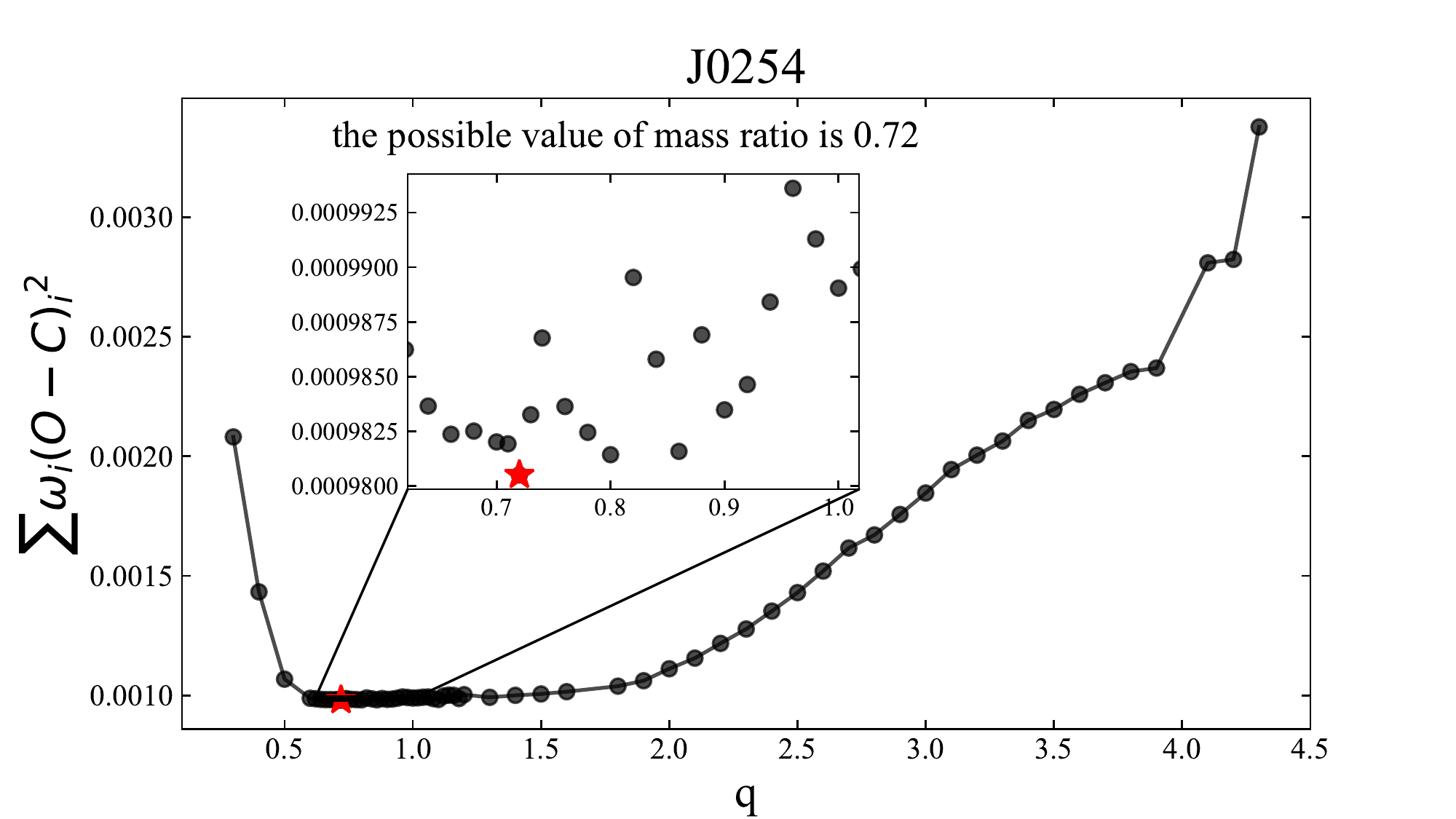}
	\includegraphics[width=0.48\textwidth]{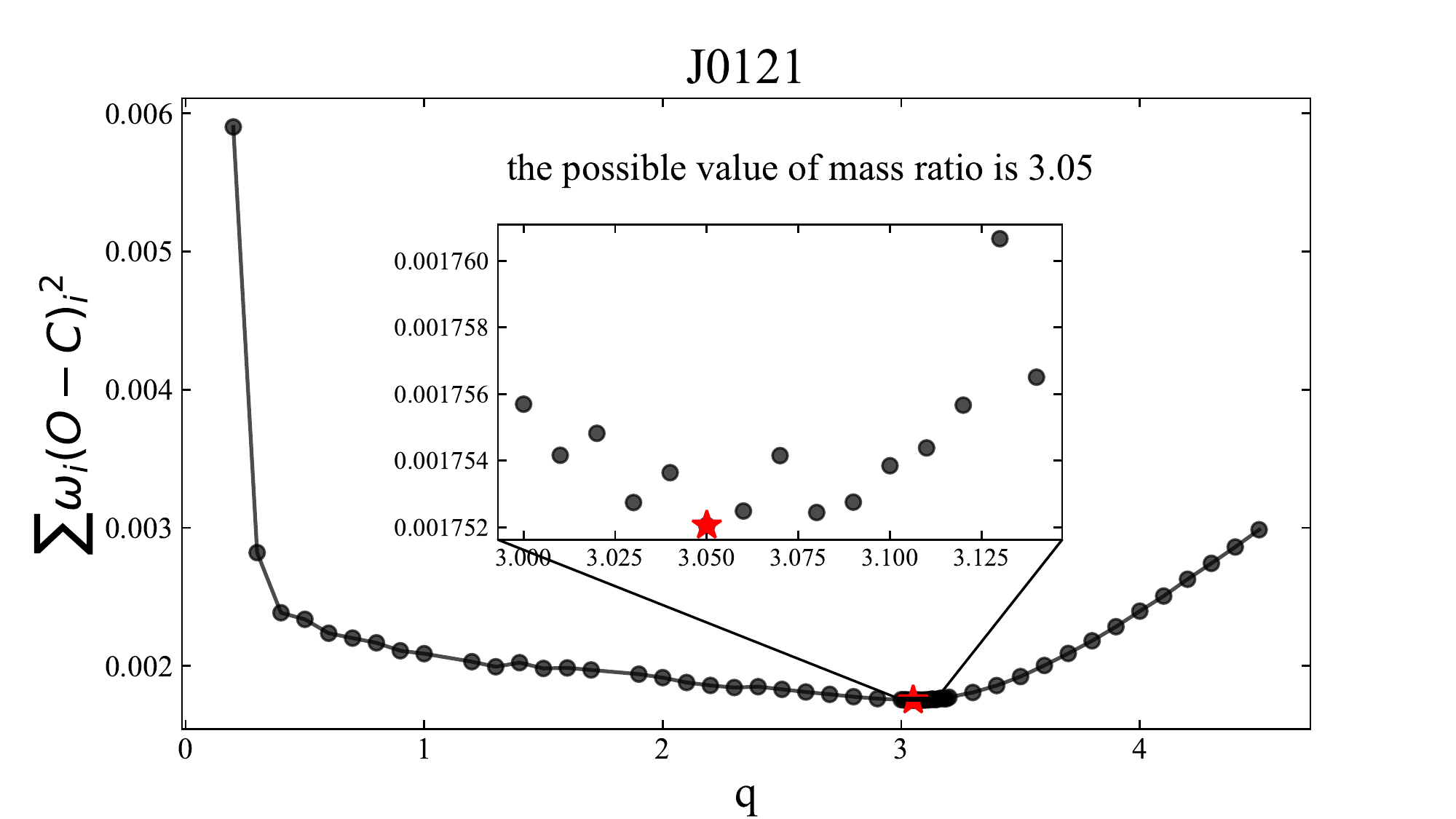}
	\caption{Relationships between sum of squares of residuals $\sum\omega_i(O - C)_i^2$ and the mass ratios $q$ for $\targeta$ and $\targetb$.}
	\label{fig:qsearch}
\end{figure*}

To further investigate $\targeta$ and $\targetb$, we analyzed our $BVRI$ bands light curves by using the 2013 version of the W-D program \citep{1971ApJ...166..605W,1979ApJ...234.1054W,1990ApJ...356..613W,2008ApJ...672..575W,2012AJ....144...73W,2010ApJ...723.1469W}.
We set the effective temperature ($T_1$) of the primary star to be the average of the spectral temperatures in Table
\ref{tab:log}, $5544 \pm 148$\,K for $\targeta$ and $5659 \pm 35 $\,K for $\targetb$. Based on this temperature, the gravity-darkening coefficients and the bolometric albedo were set to $ g_{1, 2} = 0.32 $ \citep{1967ZA.....65...89L} and $ A_{1, 2} = 0.5 $ \citep{1969AcA....19..245R} in the both targets. The bolometric and bandpass limb-darkening coefficients are estimated from \citet{1993AJ....106.2096V} with the square-root functions law. Due to the lack of radial-velocity data for the two targets, we determined the mass ratios $q$ of $\targeta$ and $\targetb$ by applying the $q$-search method to the NOWT data, and the results are shown in Fig. \ref{fig:qsearch}. We found the smallest $\Sigma\omega_i(O - C)^{2}_i$ obtained when $q=0.72$ for $\targeta$ and $q=3.05$ for $\targetb$ after using the W-D program.

When analyzing the light curve of a binary using the W-D program, we first need to determine the contact state of the two components of the binary system. But the contact states of these two targets cannot be determined yet. So we perform the calculation using mode 2, which corresponds to the detached binary. Then we found that the two components of the two targets always fill their Roche lobes during the process of operation, which means that the binaries are the contact binaries (mode 3 in the W-D program). The adjustable parameters were the orbital inclination ($i$), the mean temperature of the secondary component ($T_2$), the monochromatic luminosity of the primary component ($L_1$) in each band of $B$, $V$, $R$, $I$, and the dimensionless potential ($\Omega_1 = \Omega_2$). The photometric elements of $\targeta$ and $\targetb$ are shown in Table \ref{tab:elements}. The results show that $\targeta$ is a typical A-type W UMa contact binary, and $\targetb$ is a W-type W UMa contact binary.

\begin{figure*}[htpb]
	\centering
	\includegraphics[width=0.49\textwidth]{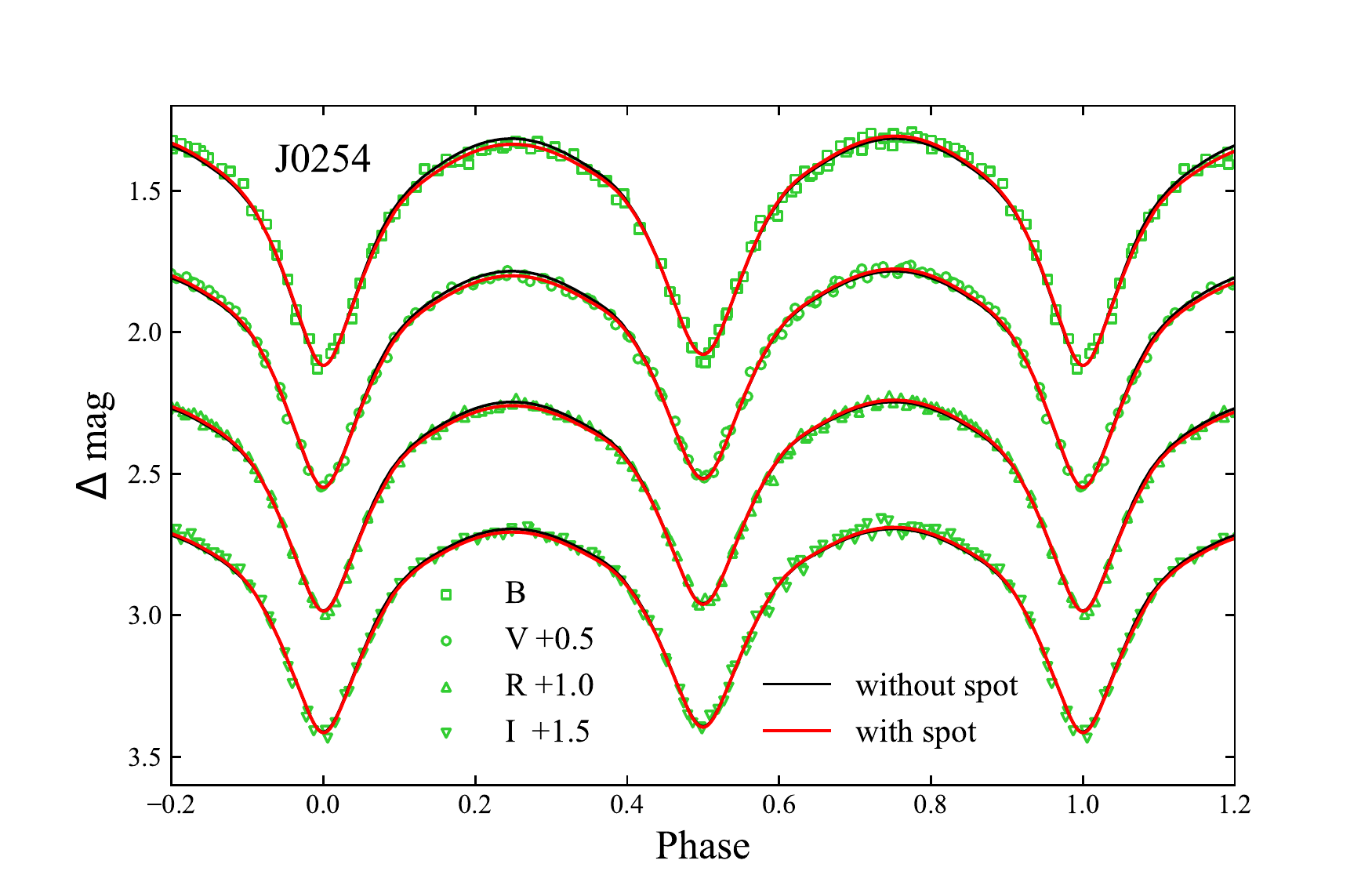}
	\includegraphics[width=0.49\textwidth]{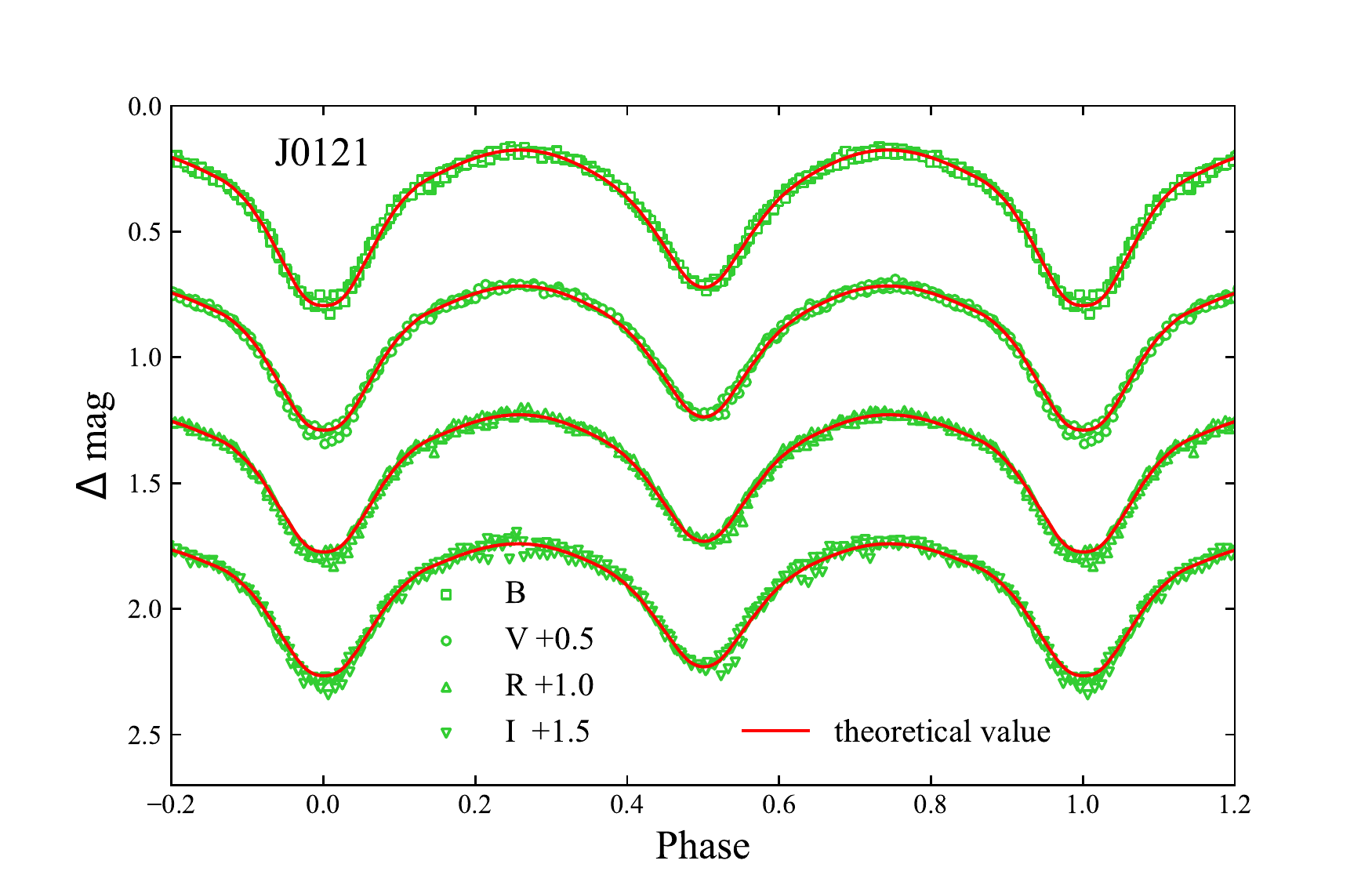}
	\caption{Theoretical and observed light curves of $\targeta$ and $\targetb$ obtained in this paper.}
	\label{fig:wdfit}
\end{figure*}

The O'Connell effect (difference in the maxima with 0.02 mag of $\targeta$ in Fig. \ref{fig:wdfit}), classically thought of as an indicator of spot activity. For $\targeta$, an excellent result can be obtained by adding a cold spot to the main component. The results are shown in Table \ref{tab:elements} and the theoretical light curves with and without spots are displayed in the left half of Fig. \ref{fig:wdfit}. We note that $ \Sigma W (O-C) ^{2} $ of 0.0017 acquired with the cool spot is smaller than $ \Sigma W (O-C) ^{2} =0.0020 $ without starspot for $\targeta$. According to the NOWT data, we found no evidence of the presence of starspot on $\targetb$.

\begin{table}[htpb]
	\footnotesize
	\caption{Photometric Elements of $\targeta$ and $\targetb$}
	\begin{tabular*}{\hsize}{@{}@{\extracolsep{\fill}}lllll@{}}
		\hline\hline
		& \multicolumn{2}{c}{$\targeta$}                             & \multicolumn{2}{c}{$\targetb$}                             \\ \hline
		Parameters                & \multicolumn{1}{c}{Values} & \multicolumn{1}{c}{Errors} & \multicolumn{1}{c}{Values} & \multicolumn{1}{c}{Errors} \\ \hline
		$ g_1=g_2 $               & 0.32                       & Assumed                    & 0.32                       & Assumed                    \\
		$ A_1=A_2 $               & 0.5                        & Assumed                    & 0.5                        & Assumed                    \\
		$ \Omega_{in} $           & 3.27848                    & Assumed                    & 6.68243                    & Assumed                    \\
		$ \Omega_{out} $          & 2.86696                    & Assumed                    & 6.06275                    & Assumed                    \\
		$ q (M_2/M_1) $           & 0.72                       & $ \pm0.001 $               & 3.05                       & $ \pm0.0022 $              \\
		$ T_1 (K) $               & 5544                       & $ \pm148.12 $              & 5659                       & $ \pm35.13 $               \\
		$ T_2 (K) $               & 5529                       & $ \pm7.73 $                & 5448                       & $ \pm3.46 $                \\
		$ i^\circ $               & 82.641                     & $ \pm0.17 $                & 79.099                     & $ \pm0.095 $               \\
		$ \Omega_{1}=\Omega_{2} $ & 3.23888                    & 0.00368                    & 6.61727                    & 0.00385                    \\
		$ L_{1B}/L_B $            & 0.5782                     & $ \pm0.0024 $              & 0.3194                     & $ \pm0.0009 $              \\
		$ L_{1V}/L_V $            & 0.5772                     & $ \pm0.0019 $              & 0.3061                     & $ \pm0.0007 $              \\
		$ L_{1R}/L_{R} $          & 0.5766                     & $ \pm0.0019 $              & 0.2982                     & $ \pm0.0006 $              \\
		$ L_{1I}/L_{I} $          & 0.5762                     & $ \pm0.0021 $              & 0.2925                     & $ \pm0.0006 $              \\
		$ r_1 \,(pole) $            & 0.3893                     & $ \pm0.0005 $              & 0.2720                     & $ \pm 0.0003 $             \\
		$ r_1 \,(side) $            & 0.4119                     & $ \pm0.0007 $              & 0.2840                     & $ \pm0.0003 $              \\
		$ r_1 \,(back) $            & 0.4442                     & $ \pm0.0009 $              & 0.3200                     & $ \pm0.0005 $              \\
		$ r_2 \,(pole) $            & 0.3348                     & $ \pm0.0005 $              & 0.4531                     & $ \pm0.0003 $              \\
		$ r_2 \,(side) $            & 0.3513                     & $ \pm0.0007 $              & 0.4870                     & $ \pm0.0003 $              \\
		$ r_2 \,(back) $            & 0.3866                     & $ \pm0.0010 $              & 0.5141                     & $ \pm0.0004 $              \\
		$ f $                     & 9.62 \%                    & $ \pm0.89 \% $             & 10.52 \%                   & $ \pm 0.62 \% $            \\
		$ \theta \,(radian) $       & 4.8184                     & $ \pm0.0690 $              &  -                         &-                            \\
		$ \phi \,(radian) $         & 1.5708                     & $ \pm0.2875 $              &  -                         &-                            \\
		$ r \,(radian) $            & 0.2203                     & $ \pm0.0080 $              &  -                         &-                            \\
		$ T_f \,(T_d/T_0) $         & 0.8196                     & $ \pm0.0179 $              &  -                         &-                            \\ \hline
	\end{tabular*}
	\label{tab:elements}
\end{table}

\begin{figure*}[htpb]
	\centering
	\includegraphics[width=0.49\textwidth]{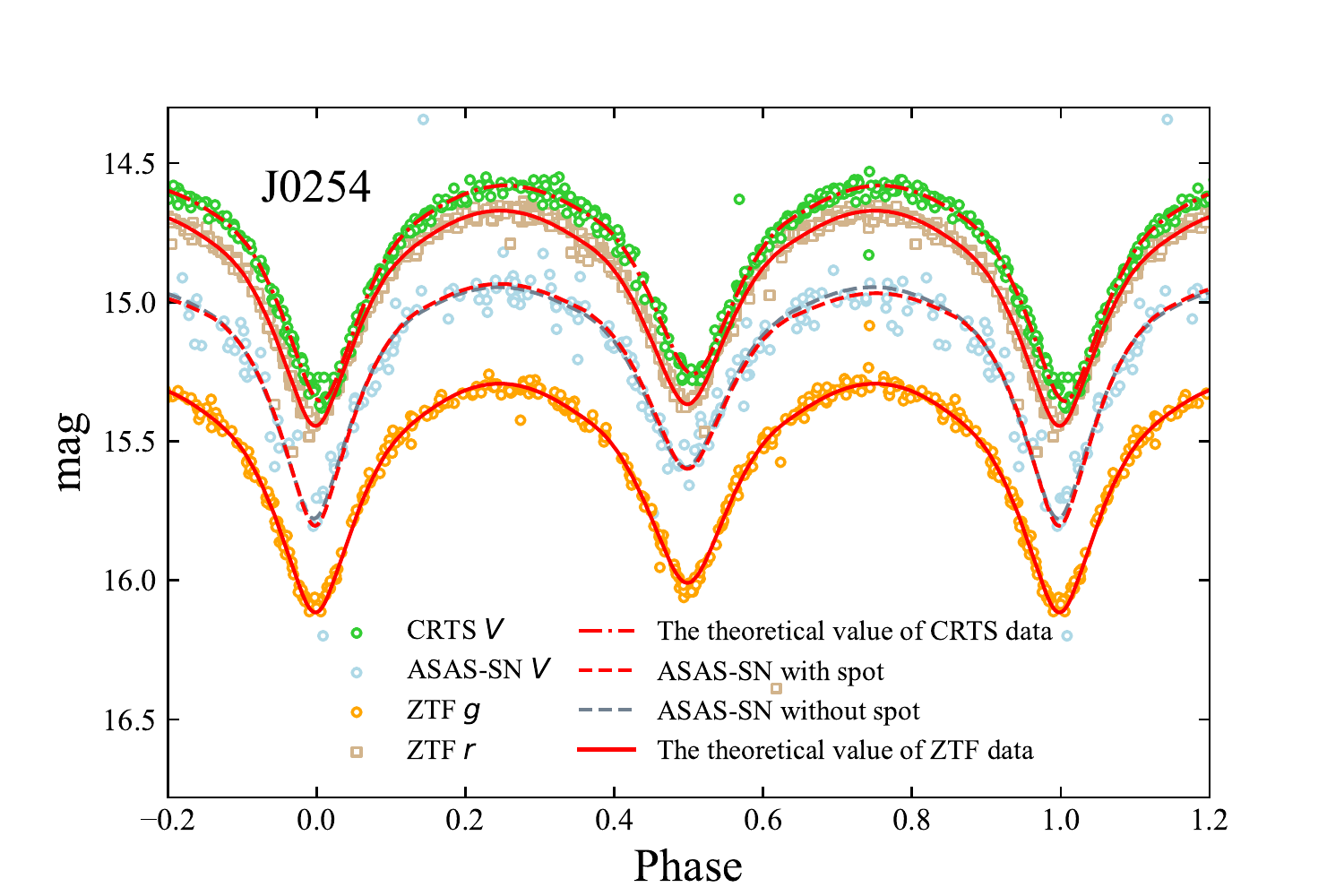}
	\includegraphics[width=0.49\textwidth]{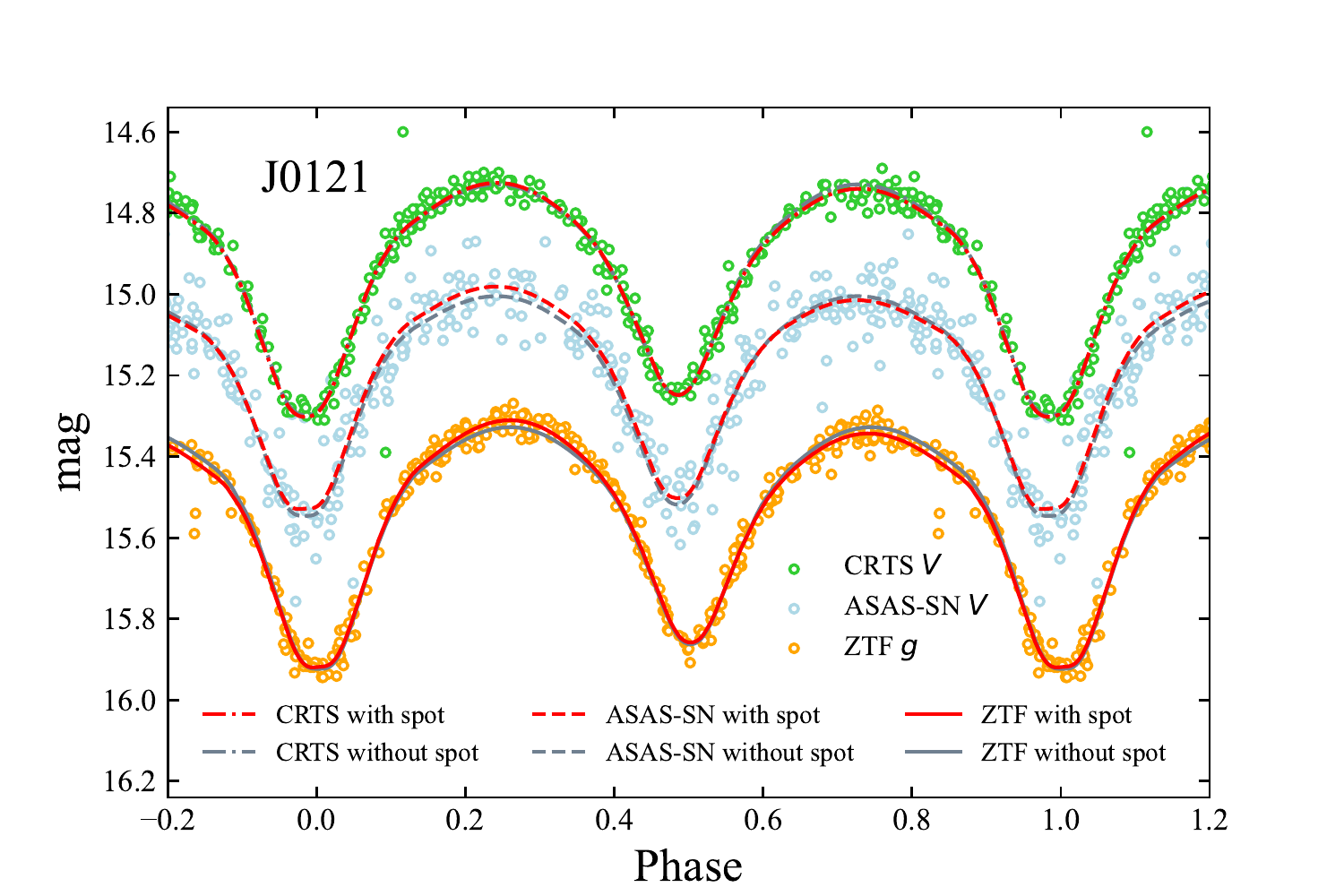}
	\caption{Theoretical and observed light curves of $\targeta$ and $\targetb$ obtained from CRTS, ASAS-SN, ZTF.}
	\label{fig:surveywdfit}
\end{figure*}

\begin{figure*}[htpb]
	\centering
	\includegraphics[width=\textwidth]{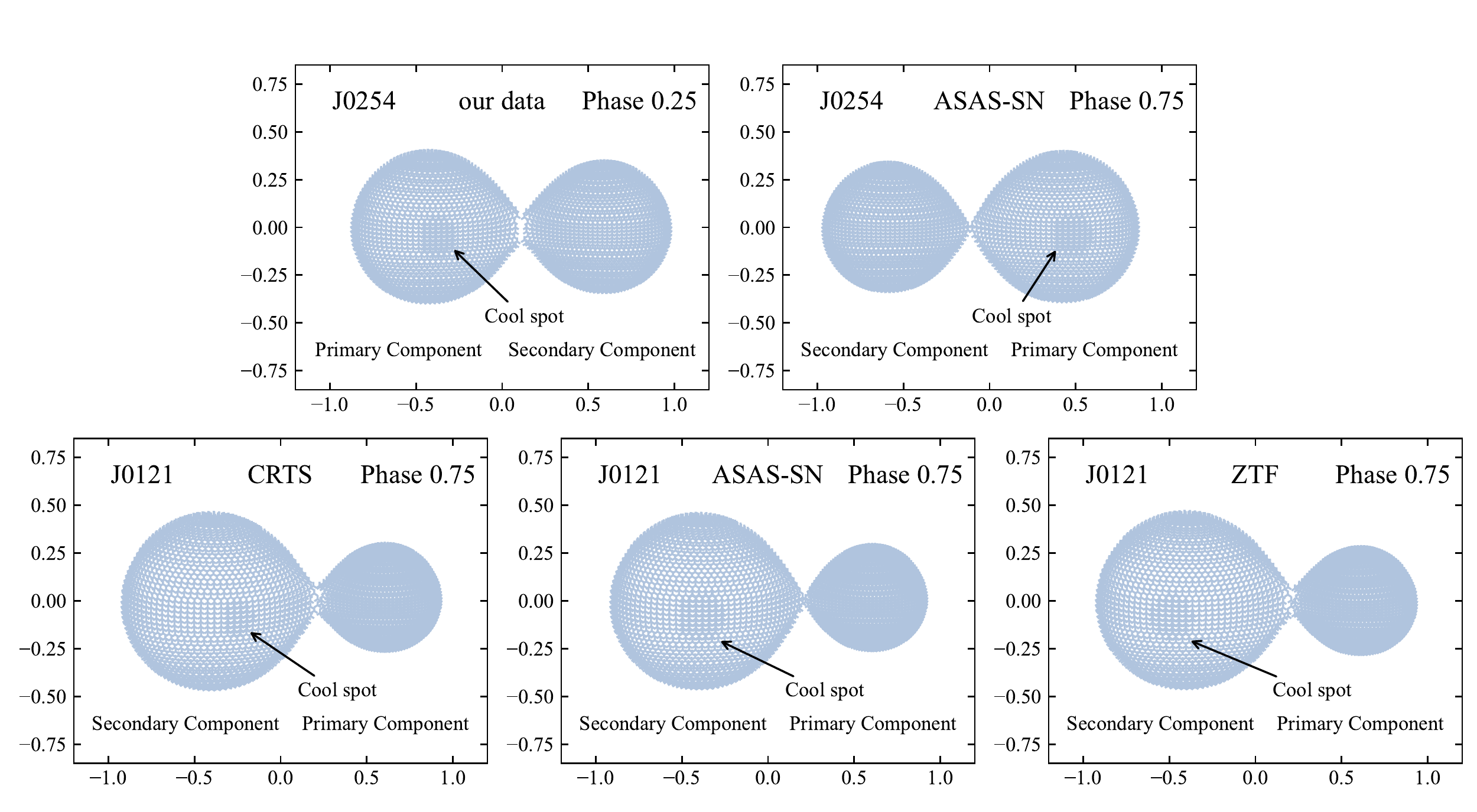}
	\caption{Stellar configuration and spot distribution for $\targeta$ and $\targetb$.}
	\label{fig:stellar}
\end{figure*}

We also performed the photometric analysis of CRTS, ASAS-SN, and ZTF data. Considering that the photometric light curve of CRTS and ASAS-SN was obtained at similar observation time of LAMOST spectra (from 2011 to 2014), we adopted the spot model to fit these light curves, as well as the same input parameters mentioned above. When running the program, in order to simplify the model, the spot latitude $\phi$ is fixed at $90^{\circ}$ (1.5708 $radian$) and the temperature factor $ T_f \,(T_d/T_0) $ is fixed at 0.8. We found the presence of a cold spot on the more massive component of $\targeta$ based on the ASAS-SN data, but no starspot is present in the CRTS with ZTF data results. While the results of $\targetb$ in CRTS, ASAS-SN, and ZTF data all indicate the presence of a cold spot on its massive component. The optimal results are presented in Table \ref{tab:elementssurvey}, and the fitted theoretical light curves are plotted in Fig. \ref{fig:surveywdfit}. The corresponding stellar structure is presented in Fig. \ref{fig:stellar}.

\begin{table*}[htpb]
	\tiny
	\caption{Photometric Elements of $\targeta$ and $\targetb$ based on survey data}
	\begin{tabular*}{\hsize}{@{}@{\extracolsep{\fill}}lcccccc@{}}
		\hline\hline
		& \multicolumn{3}{c}{$\targeta$}                             & \multicolumn{3}{c}{$\targetb$}                             \\ \hline
		Parameters                & \multicolumn{1}{c}{CRTS} & \multicolumn{1}{c}{ASAS-SN} & \multicolumn{1}{c}{ZTF}   & \multicolumn{1}{c}{CRTS} & \multicolumn{1}{c}{ASAS-SN} & \multicolumn{1}{c}{ZTF}  \\ \hline
		$ q (M_2/M_1) $           & 0.72                      & 0.72                        &0.72                        & 3.05                      & 3.05                         &  3.05                    \\
		$ T_1 (K) $               & 5544                      & 5544                        &5544                        & 5659                      & 5659                         &  5659                    \\
		$ T_2 (K) $               & 5410       $\pm$ 17.59    & 5287       $\pm$ 94.10      & 5430     $\pm$ 18.42       & 5436     $\pm$ 35.71      & 5498     $\pm$ 64.85         & 5442     $\pm$ 16.38      \\
		$ i^\circ $               & 81.859     $\pm$ 0.267    & 84.044     $\pm$ 1.707       & 82.345   $\pm$ 0.304       & 79.028   $\pm$ 0.627      & 80.162   $\pm$ 1.573         & 79.953   $\pm$ 0.442      \\
		$ \Omega_{1}=\Omega_{2} $ & 3.2486    $\pm$ 0.0078  & 3.2776    $\pm$ 0.0303    & 3.2268  $\pm$ 0.0080     & 6.6178  $\pm$ 0.0253    & 6.6715  $\pm$ 0.0479       & 6.6115  $\pm$ 0.0167    \\
		$ L_{1V}/L_V $            & 0.6049     $\pm$ 0.0041   & 0.5772     $\pm$ 0.0231     &  -                         & 0.3086   $\pm$ 0.0072     & 0.2937   $\pm$ 0.0125        &   -                      \\
		$ L_{1g}/L_{g} $          & -                         &  -                          &0.6037     $\pm$ 0.0053     & -                         &  -                           & 0.3148   $\pm$ 0.0040     \\
		$ L_{1r}/L_{r} $          & -                         &  -                          &0.5960     $\pm$ 0.0038     & -                         &  -                           &   -                      \\
		$ r_1 (pole) $            & 0.3880     $\pm$ 0.0010   & 0.3835     $\pm$ 0.0043     &0.3913     $\pm$ 0.0012     & 0.2724   $\pm$ 0.0018     & 0.2676   $\pm$ 0.0033        & 0.2725   $\pm$ 0.0012     \\
		$ r_1 (side) $            & 0.4102     $\pm$ 0.0014   & 0.4046     $\pm$ 0.0054     &0.4143     $\pm$ 0.0015     & 0.2844   $\pm$ 0.0021     & 0.2786   $\pm$ 0.0038        & 0.2845   $\pm$ 0.0014     \\
		$ r_1 (back) $            & 0.4418     $\pm$ 0.0020   & 0.4342     $\pm$ 0.0072     &0.4475     $\pm$ 0.0020     & 0.3207   $\pm$ 0.0035     & 0.3112   $\pm$ 0.0061        & 0.3209   $\pm$ 0.0024     \\
		$ r_2 (pole) $            & 0.3333     $\pm$ 0.0012   & 0.3291     $\pm$ 0.0044     &0.3366     $\pm$ 0.0012     & 0.4531   $\pm$ 0.0017     & 0.4496   $\pm$ 0.0031        & 0.4535   $\pm$ 0.0011     \\
		$ r_2 (side) $            & 0.3496     $\pm$ 0.0014   & 0.3445     $\pm$ 0.0053     &0.3535     $\pm$ 0.0015     & 0.4869   $\pm$ 0.0022     & 0.4822   $\pm$ 0.0042        & 0.4875   $\pm$ 0.0015     \\
		$ r_2 (back) $            & 0.3839     $\pm$ 0.0021   & 0.3764     $\pm$ 0.0076     &0.3899     $\pm$ 0.0022     & 0.5141   $\pm$ 0.0028     & 0.5081   $\pm$ 0.0052        & 0.5148   $\pm$ 0.0019     \\
		$spot$                    &            -              & primary star                &            -               &   secondry star           &   secondry star              &   secondry star          \\
		$ f $                     & 5.29 \%    $\pm$ 1.90 \%  & 0.22 \%    $\pm$ 7.36 \%     &12.55 \%    $\pm$ 1.94 \%   & 10.43 \  $\pm$  4.08 \%   & 1.76 \  $\pm$  7.72 \%       & 11.45 \  $\pm$  2.70 \%   \\
		$ \theta (radian) $       &            -              & 1.7078    $\pm$ 0.6222    &           -                & 4.9995     $\pm$ 0.5052   &4.6397     $\pm$ 0.3150       & 4.5441   $\pm$ 0.0903    \\
		$ \phi (radian) $         &            -              & 1.5708                      &           -                & 1.5708                    &1.5708                        & 1.5708                   \\
		$ r (radian) $            &            -              & 0.2380    $\pm$ 0.0678    &           -                & 0.1491     $\pm$ 0.0336   &0.2233     $\pm$ 0.0325       & 0.2146   $\pm$ 0.0115     \\
		$ T_f (T_d/T_0) $         &            -              & 0.8                         &           -                & 0.8                       &0.8                           & 0.8                      \\ \hline
	\end{tabular*}
	\label{tab:elementssurvey}
\end{table*}

As shown in the above results, from the CRTS and ZTF data, no starspot was found on $\targeta$. This may be due to the fact that the two components of $\targeta$ have similar masses, so that the chances of magnetic activity occurring in either component may be equal, which makes the dispersion increase at the two maxima of the light curve, thus masking the starspot. In conclusion, the photometric solutions for $\targeta$ and $\targetb$ indicate the presence of starspots. Combined with our chromospheric activity analysis, it is confirmed that the two targets have magnetic activity.

\section{Discussion} \label{sec:discussions}

In this section, we discussed absolution parameters estimation with different methods (Section \ref{subsec:abspar}), calculated the distances of the two targets (Section \ref{subsec:distance}), and analyzed the evolution for our targets (Section \ref{subsec:transfer and evolution}).

\subsection{Absolution parameters} \label{subsec:abspar}

The absolute parameters of binary usually include mass $M$, radius $R$, luminosity $L$, and the semi-major axis of the orbit $a$. These parameters are linked to each other. In a binary system, the accurate absolute parameters can be determined in the case of having radial-velocity curves. However, in the absence of the radial-velocity curves, we can only some estimation methods to assess the absolute parameters of the binary system.

\citet{2017MNRAS.466.1118Z} proposed a new method for estimating absolute parameters and provided mass-radius relations of two contact binary systems (GQ Boo and V1367 Tau). They obtained the approximate masses and radii by using the relations to limit the almost complete stellar parameter space provided by PARSEC\footnote{PAdova and Trieste Stellar Evolution Code \url{http://stev.oapd.inaf.it/cgi-bin/cmd}} \citep{2012MNRAS.427..127B}, which include all the initial stellar masses (0.1$ <$ $M$(M$_{\odot}$) $<$ 350) and ages (6.6 $<$ $\log (t$/yr) $<$ 10.13), the metallicity Z = 0.010-0.023 are selected for GQ Boo, and Z = 0.0001-0.0700 for V1367 Tau. They pointed out that the more massive components are more suitable than the less massive components for matching the absolute parameters, with the temperature of 200K error being used to limit the more massive components. They also pointed out that single stars are brighter and hotter than components of the same mass in a contact binary system, that there may be some bias in the estimates using the single-star evolution procedure, and that the PARSEC program may be more applicable to detached binaries. The method is used to contact binaries resulting in biases that are within the errors of the absolute parameters obtained. The setting of the atmospheric input parameters has the most influence on this method.

\citet{2019RAA....19..108W} provided a method depending on the calculation of the Roche lobes (the DCRL method), while referring to the method used by \citet{2017MNRAS.466.1118Z} as DCWD, which depends on the stellar radius obtained from the W-D code. In their paper, these two methods are used to obtain the absolute parameters of AL Cas.

\citet{2020RAA....20..196L} provided a general formula for the DCRL method suitable for any contact binary system. The DCRL method assumes that the radius of the companion star coincides with the effective radius of the Roche lobe $R_{L}$ that can be obtained by the following relation \citep{1983ApJ...268..368E}:
\begin{equation}
	\frac{R_{L}}{a}= \frac{0.49q^{2/3}}{0.6q^{2/3}+ \ln(1+q^{1/3})},
	\label{eq:rocheradius}
\end{equation}
where $q$ is mass radio, $q = M_1/M_2$ for $R_{L_1}$ and $q = M_2/M_1$ for $R_{L_2}$.

According to Equation (\ref{eq:rocheradius}), the mass-radius relation of DCRL was derived as:
\begin{equation}
	\frac{R}{{R}_{\odot}}=2.0627\left[\frac{q^{1 / 3} p^{2 / 3}(q+1)^{1 / 3}}{0.6 q^{2 / 3}+ln\left(1+q^{1 / 3}\right)}\right]\left(\frac{M}{M_{\odot}}\right)^{1 / 3},
	\label{eq:DCRL}
\end{equation}
where $0 < q < 1$, and $P$ is the orbital period. The star parameter space is obtained by PARSEC v1.2S stellar evolution code, and the parameters are listed in Table \ref{tab:spacePARSEC}. They tested 140 binary systems (76 W-type and 64 A-type systems) with sufficient spectral information and found that their method is suitable for short-period W UMa systems with a high mass ratio and low effective temperature. Since the fractional difference $\left(\frac{{Mass}_{A v=0}-{Mass}_{A v \neq 0}}{{Mass} A v=0}\right)$ for all sets of parameters is found less than 2.4\%, A$_v$ was set as 0 by them.

\begin{table}[htbp]
	\begin{center}
	\centering
	\caption{Parameter space of PARSEC.}
	\begin{tabular}{lcc}
		\hline\hline
		Name                 &  Parameter space & Reference                \\ \hline
		Evolutionary tracks  & PARSEC version 1.2S     & 1,2,3   \\
		Photometric systems  & $U B V R I J H K$       & 4,5,6 \\
		Circumstellar dust   &  No dust  &         7 \\
		Mass                 & 0.1 M$_{\odot}<M<$350 M$_{\odot}$ & 1,2 \\
		Age                  & $6.6<\log (t)<10.13$ & 1,2 \\
		Metallicities        & $0.0001<Z<0.07$ & 1,2 \\
		Interstellar extinction & $R_{v}=3.1, A_{v}=0$ & 8,9 \\ \hline
	\end{tabular}
	\label{tab:spacePARSEC}
	\end{center}
	\tablecomments{0.48\textwidth}{1 \citet{2014MNRAS.445.4287T}; 2 \citet{2015MNRAS.452.1068C}; 3 \citet{2014MNRAS.444.2525C}; 4 \citet{2006AJ....131.1184M}; 5 \citet{1990PASP..102.1181B}; 6 \citet{1988PASP..100.1134B}; 7 \citet{2008A&A...482..883M}; 8 \citet{1989ApJ...345..245C}; 9 \citet{1994ApJ...422..158O}.}
\end{table}

In this paper, we provide a generic formulation for the DCWD method that can be used for any contact binary and requires only some of the parameters of that binary system. Combining Equation (\ref{eq:rocheradius}) with Kepler's third law, we obtain the mass-radius relationship:
\begin{equation}
	\frac{R}{{R}_{\odot}}=\frac{2.4089(1+{q})^{1 / 3} P^{2 / 3} {r}}{{q}^{1 / 3}}\left(\frac{M}{M_{\odot}}\right)^{1 / 3},
	\label{eq:DCWD}
\end{equation}
where $0 < q < 1$, and $r$ is the result of the ratio of the companion radius to the semi-major axis of orbit obtained by the photometric analysis using the W-D program.

The mass-radius relations $\targeta$ and $\targetb$ are obtained from the orbital period $P$, the mass ratio $q$, the ratio of radius and semi-major axis $r= R/a$ and Equations (\ref{eq:DCRL}) and (\ref{eq:DCWD}).\\
For $\targeta$,
\begin{eqnarray}
	\frac{M_1}{M_{\odot}} & = & 0.84550(\pm0.00027)\left(\frac{R_1}{R_{\odot}}\right)^{3}_{DCRL}\text{,} \\ \nonumber
	\frac{M_1}{M_{\odot}} & = & 0.88897(\pm0.00355)\left(\frac{R_1}{R_{\odot}}\right)^{3}_{DCWD}\text{.}
\end{eqnarray}
For $\targetb$,
\begin{eqnarray}
	\frac{M_2}{M_{\odot}} & = & 1.12407(\pm0.00025)\left(\frac{R_2}{R_{\odot}}\right)^{3}_{DCRL}\text{,} \\ \nonumber
	\frac{M_2}{M_{\odot}} & = & 1.17083(\pm0.00443)\left(\frac{R_2}{R_{\odot}}\right)^{3}_{DCWD}\text{.}
\end{eqnarray}

In this work, we use the PARSEC program of CMD version 3.6. For its input parameter space, there are three considerations.
\begin{enumerate}
	\item In case 1, the complete parameter space without restriction. The circumstellar dust is chosen for "No dust" mode, all the initial stellar masses (0.1$ <$ $M$ (M$_{\odot}$) $<$ 350) are used. The evolution model settings use the default values. The metallicity $Z$ from 0.0001 to 0.06 in steps of 0.0001 and the logarithmic age from 6.6 to 10.13 in steps of 0.05.
	\item In case 2, since \cite{2014MNRAS.437..185Y} indicated that the average age of W UMa systems is around 4.4-4.6 Gyr, it varies depending on type A or W, so we limit the input to ages above $10^8$. The age with logarithmic from 8 to 10.13 in steps of 0.05, other input parameters are the same as in case 1.
	\item In case 3, the logarithmic age from 8 to 10.13 in steps of 0.01, and the step of metallicity Z is 0.0001. Here, $0.0063<Z<0.0170$ for $\targeta$ and $0.0080<Z<0.0109$ for $\targetb$ are used, and the metallicity is derived from low-resolution spectra of LAMOST. Other input parameters are the same as in case 1.
\end{enumerate}

\begin{figure*}[htpb]
	\centering
	\includegraphics[width=0.49\textwidth]{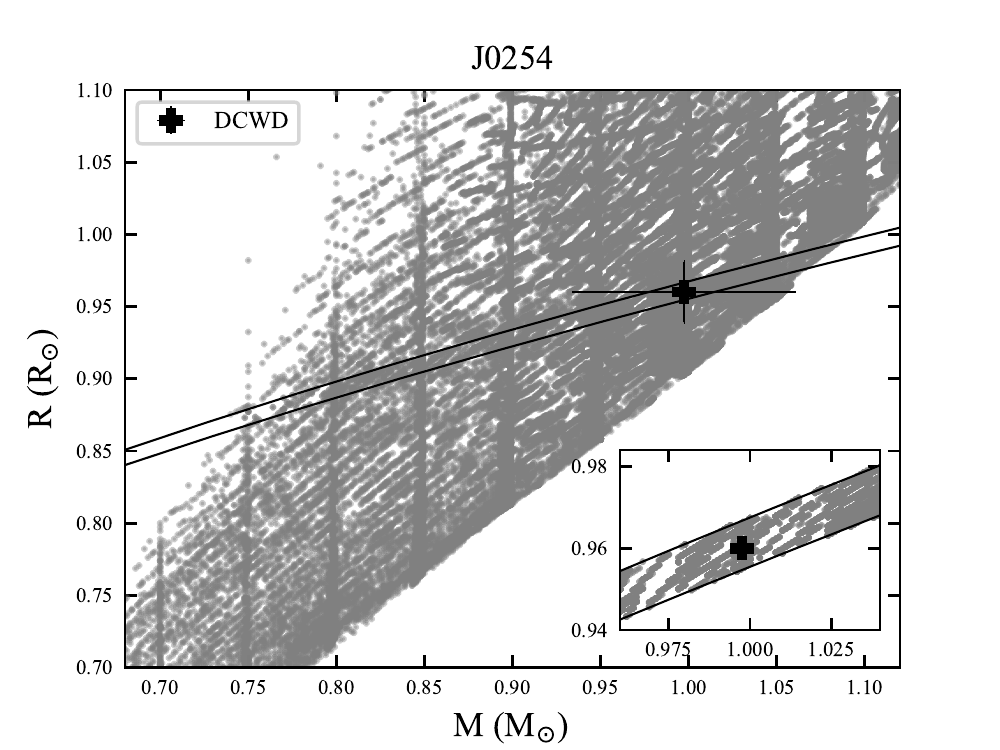}
	\includegraphics[width=0.49\textwidth]{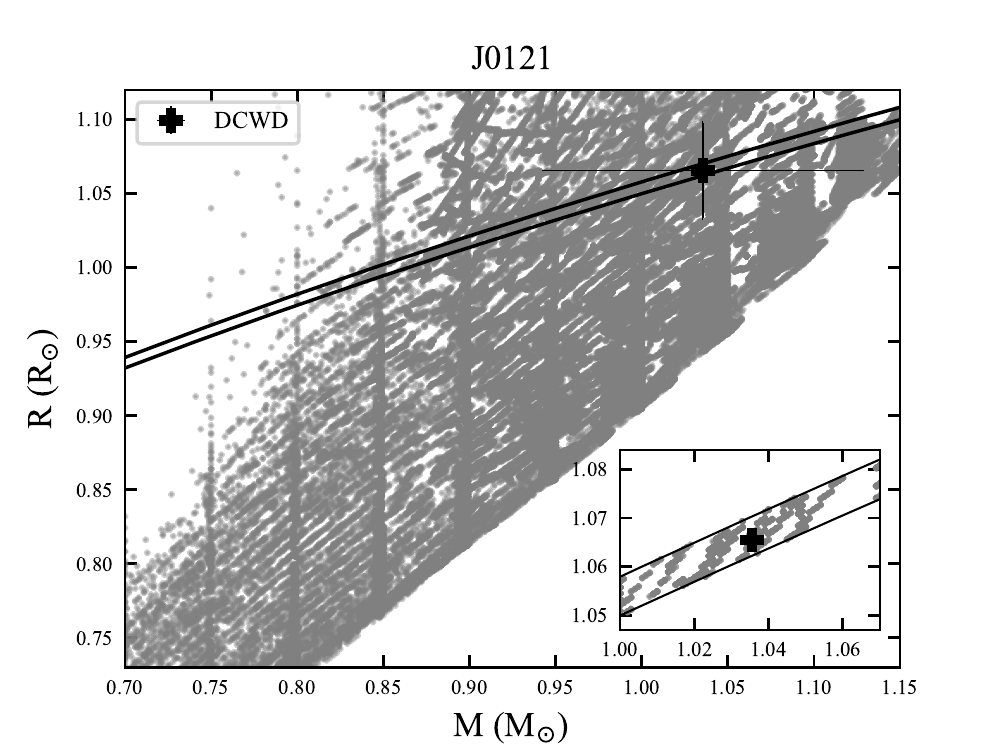}
	\caption{The stellar M-R relation of $\targeta$ and $\targetb$ using the DCWD method in case 1.}
	\label{fig:DCWD}
\end{figure*}

We floated the temperature at 300 K to constrain the PARSEC output parameters. Using the DCWD method in case 1 as an example, the results of the two targets are shown in Fig. \ref{fig:DCWD}. Data within a triple error margin of the mass-radius relationship for both targets were used to estimate the approximate absolute parameters (mass and radius) for the more massive component. We also calculated other absolute parameters and all results are presented in Table \ref{tab:absdcrldcwd}.

\begin{table*}[htpb]
	\centering
	\tiny
	\caption{The absolute parameters of $\targeta$ and $\targetb$ are obtained by using DCRL and DCWD methods.}
	\begin{tabular*}{\hsize}{@{}@{\extracolsep{\fill}}llcccccc@{}}
		\hline\hline
		
		Object           &		Model  & \multicolumn{3}{c}{DCRL}   & \multicolumn{3}{c}{DCWD}         \\ \hline
		      &Parameters & case 1 & case 2 &case 3   & case 1 & case 2 &case 3  \\ \cline{2-8}
&$M_1$(M$_\odot$)&0.988$\pm$0.068&0.963$\pm$0.062&0.900$\pm$0.058&0.998$\pm$0.063&0.997$\pm$0.059&0.909$\pm$0.062\\
&$M_2$(M$_\odot$)&0.711$\pm$0.049&0.693$\pm$0.045&0.648$\pm$0.042&0.718$\pm$0.046&0.718$\pm$0.044&0.654$\pm$0.046\\
&$R_1$(R$_\odot$)&0.941$\pm$0.022&0.933$\pm$0.021&0.913$\pm$0.020&0.960$\pm$0.021&0.960$\pm$0.020&0.931$\pm$0.022\\
$\targeta$&$R_2$(R$_\odot$)&0.810$\pm$0.250&0.803$\pm$0.247&0.785$\pm$0.241&0.827$\pm$0.021&0.827$\pm$0.020&0.801$\pm$0.022\\
&$a$(R$_\odot$)    &2.309$\pm$0.394&2.289$\pm$0.388&2.239$\pm$0.378&2.316$\pm$0.054&2.316$\pm$0.052&2.245$\pm$0.056\\
&$L_1$(L$_\odot$)&0.750$\pm$0.115&0.737$\pm$0.112&0.705$\pm$0.106&0.780$\pm$0.117&0.780$\pm$0.116&0.733$\pm$0.112\\
&$L_2$(L$_\odot$)&0.549$\pm$0.342&0.540$\pm$0.335&0.517$\pm$0.320&0.572$\pm$0.033&0.572$\pm$0.031&0.538$\pm$0.032\\ \cline{2-8}
&$M_1$(M$_\odot$)&0.335$\pm$0.026&0.336$\pm$0.026&0.281$\pm$0.003&0.339$\pm$0.031&0.330$\pm$0.027&0.281$\pm$0.007\\
&$M_2$(M$_\odot$)&1.020$\pm$0.079&1.024$\pm$0.079&0.856$\pm$0.010&1.035$\pm$0.092&1.006$\pm$0.081&0.858$\pm$0.021\\
&$R_1$(R$_\odot$)&0.630$\pm$0.166&0.631$\pm$0.166&0.595$\pm$0.143&0.641$\pm$0.021&0.635$\pm$0.018&0.603$\pm$0.006\\
$\targetb$&$R_2$(R$_\odot$)&1.046$\pm$0.027&1.047$\pm$0.027&0.987$\pm$0.004&1.066$\pm$0.032&1.055$\pm$0.029&1.001$\pm$0.009\\
&$a$(R$_\odot$)    &2.191$\pm$0.346&2.193$\pm$0.346&2.067$\pm$0.280&2.201$\pm$0.068&2.180$\pm$0.061&2.069$\pm$0.020\\
&$L_1$(L$_\odot$)&0.365$\pm$0.201&0.366$\pm$0.201&0.325$\pm$0.164&0.378$\pm$0.034&0.371$\pm$0.031&0.334$\pm$0.015\\
&$L_2$(L$_\odot$)&0.864$\pm$0.047&0.866$\pm$0.047&0.769$\pm$0.008&0.896$\pm$0.056&0.879$\pm$0.050&0.792$\pm$0.016\\ \hline
	\end{tabular*}
	\label{tab:absdcrldcwd}
\end{table*}

From the results, when the parameter input space of the PARSEC program was constrained by age, the maximum range of variation of the absolute parameters was 2.8\%. The parameters of the two targets have different responses for the age constraint in case 1 and case 2. From the assumptions of case 1 and case 2, the age space changes from  6.6 $<$ $\log (t$/yr) $<$ 10.13 to 8.0 $<$ $\log (t$/yr) $<$ 10.13. and the estimated parameters of $\targeta$ decrease while those of $\targetb$ remain the same or even increase slightly when using the DCRL method. In contrast, when using the DCWD method, the estimated parameters of $\targeta$ are unchanged and those of $\targetb$ are decreased. This situation may be due to the difference in the mass ratio of the two targets, or the difference in the models, which may require more target samples to analyze the causes. The results change significantly from case 2 to case 3, where we overestimate the absolute parameters of the targets by about 10-15\% without considering metallicity, which may be caused by the reduction of the parameter space making very few data satisfying the mass-radius relationship. The absolute parameter estimation method combined with the PARSEC stellar evolution model would underestimate the parameters of the target, since the more massive part of the binary is fainter and cooler than a single star with the same mass. Thus, it is closer to the actual value without restricting the atmospheric parameters. Combining the radial velocity profiles could yield physical parameters that approximate the actual values. Therefore, it is necessary to obtain radial velocity profiles by additional spectroscopic observations.


\subsection{Distances} \label{subsec:distance}

Using the DCRL with DCWD method, the obtained luminosity of the more massive component of our targets $\targeta$ and $\targetb$ are applied to the calculation of distances. The formula used for this calculation is as follows: (a) $-2.5 \log (L / L_{\odot})=M_{{bol}}-4.73$ \citep{2010AJ....140.1158T}, (b) ${M}_{bol _i}={M}_{V_i}+{BC}_{V_i}$, (c) ${m}_{V_i}-{m}_{V _{\max} }=-2.5 \log \left({L}_{{i}} / {L}\right)$, (d) ${M}_{V_{i}}={m}_{V_{i}}-5 \log {D}+5-{A}_{V}$. During the calculation, some parameters are necessary, including the interstellar extinction coefficient $A_{v_{S \& F}}$ \citep{2011ApJ...737..103S} from IRSA database\footnote{\url{https://irsa.ipac.caltech.edu/applications/DUST/}} ($A_v = 0.367$ for $\targeta$ and $A_v = 0.226$ for $\targetb$), the bolometric corrections from \citet{2013ApJS..208....9P} ($BC_V=-0.126$ for $\targeta$ and $BC_V=-0.146$ for $\targetb$), and the maximum visual magnitude $m_{V_{\max}}$ obtained by fitting ASAS-SN data ($m_{V_{\max}}=14.945$ mag for $\targeta$ and $m_{V_{\max}}=15.005$ mag for $\targeta$). The results are listed in Table \ref{tab:distance}. In order to verify the reliability of the parameter estimation method we used and the distances obtained, we used the distances provided by \citet{2018AJ....156...58B} (hereafter BJ18), the photometric distances provided by \citet{2021AJ....161..147B} (hereafter BJ21) and the distances obtained from Gaia DR3 \citep{2022yCat.1355....0G} for comparison with the distances we obtained.

\begin{table}[htpb]
	\begin{center}
	\caption{Distances of $\targeta$ and $\targetb$.}
	\begin{tabular}{lcc}
		\hline\hline
		& \multicolumn{2}{c}{Distance(pc)} \\
		Source &$\targeta$ & $\targetb$\\ \hline
		BJ18$^1$&1093 $\pm$ 49  &1177 $\pm$ 56\\
		BJ21$^2$&1045 $\pm$ 38  & 1205 $\pm$ 47\\
		Gaia DR3$^3$ & 1091 $\pm$ 34 & 1149 $\pm$ 44\\
		DCRL $_{case\,1}$    & 1003 $\pm$ 135 & 1067 $\pm$ 101 \\
		DCRL $_{case\,2}$    & 994  $\pm$ 133 & 1068 $\pm$ 101 \\
		DCRL $_{case\,3}$    & 973  $\pm$ 130 & 1007 $\pm$ 72  \\
		DCWD $_{case\,1}$    & 1023 $\pm$ 136 & 1087 $\pm$ 107 \\
		DCWD $_{case\,2}$    & 1023 $\pm$ 135 & 1076 $\pm$ 103 \\
		DCWD $_{case\,3}$    & 992  $\pm$ 133 & 1022 $\pm$ 79  \\
		\hline
	\end{tabular}
	\label{tab:distance}
	\end{center}
	\tablecomments{0.48\textwidth}{1 \citet{2018AJ....156...58B}; 2 \citet{2021AJ....161..147B}; 3 \citet{2022yCat.1355....0G}.}
\end{table}
By comparison, we find that the obtained distances are in general agreement with the distances in the literature, which proves the validity of the method we adopted. From the results, the parameters obtained in case 1 may be closer to the actual values.

\subsection{Mass transfer and evolution of angular momentum} \label{subsec:transfer and evolution}

In this section, we use the absolute parameters obtained by the DCWD method in the case 1 condition (see table \ref{tab:absdcrldcwd}) to study the evolution of the two targets.

The $(O-C)$ analysis shows that for both $\targeta$ and $\targetb$, there is a long-term trend of increasing orbital periods for both systems. This phenomenon can be explained by mass transfer. The equation ($ \frac{\dot{P}}{P}=3 \dot{M}_{1}\left(\frac{1}{M_{2}}-\frac{1}{M_{1}}\right) $) given by \citet{1991mnras.253....9T} was used to calculate the mass transfer rate of binary. The mass transfer rates of our two targets were obtained, with $\targeta$ ($d M_{1} / d t= 2.682 \times 10^{-7} \rm{M_{\odot}}$ yr$^{-1}$) and $\targetb$ ($d M_{1} / d t= -3.603 \times 10^{-7} \rm{M_{\odot}}$ yr$^{-1}$). From the mass transfer rates and the mass ratio of the two targets, we suggested that both targets are transferring mass from the less massive components to the more massive components.
This means that these two targets are potentially excellent targets for proving the theory of thermal relaxation oscillations \citep{1967ZA.....65...89L,1979ApJ...231..502L}.
Mass transfer from lower to higher mass components in the W UMa system leads to an increase in the orbital angular momentum. According to the thermal relaxation oscillation model, these two targets may evolve as semi-detached binaries.
Our two targets have magnetic activity, so the possibility that magnetic activity plays a role in the orbital period variation cannot be denied. However, due to the insufficient accumulation of occultation times, further studies of the orbital period variation of these two targets are needed in the future.

\begin{figure*}[htbp]
	\centering
	\includegraphics[width=0.49\textwidth]{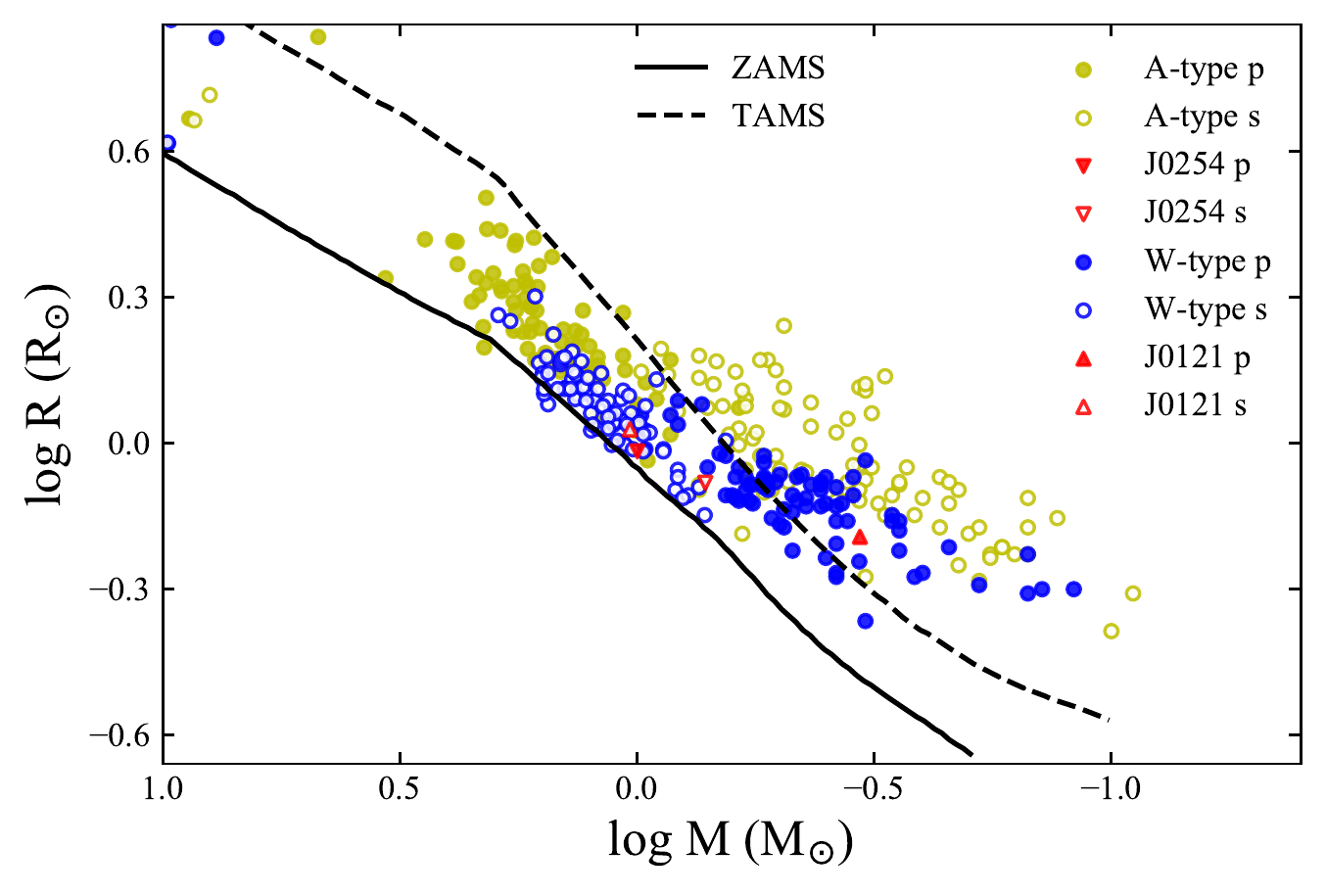}
	\includegraphics[width=0.49\textwidth]{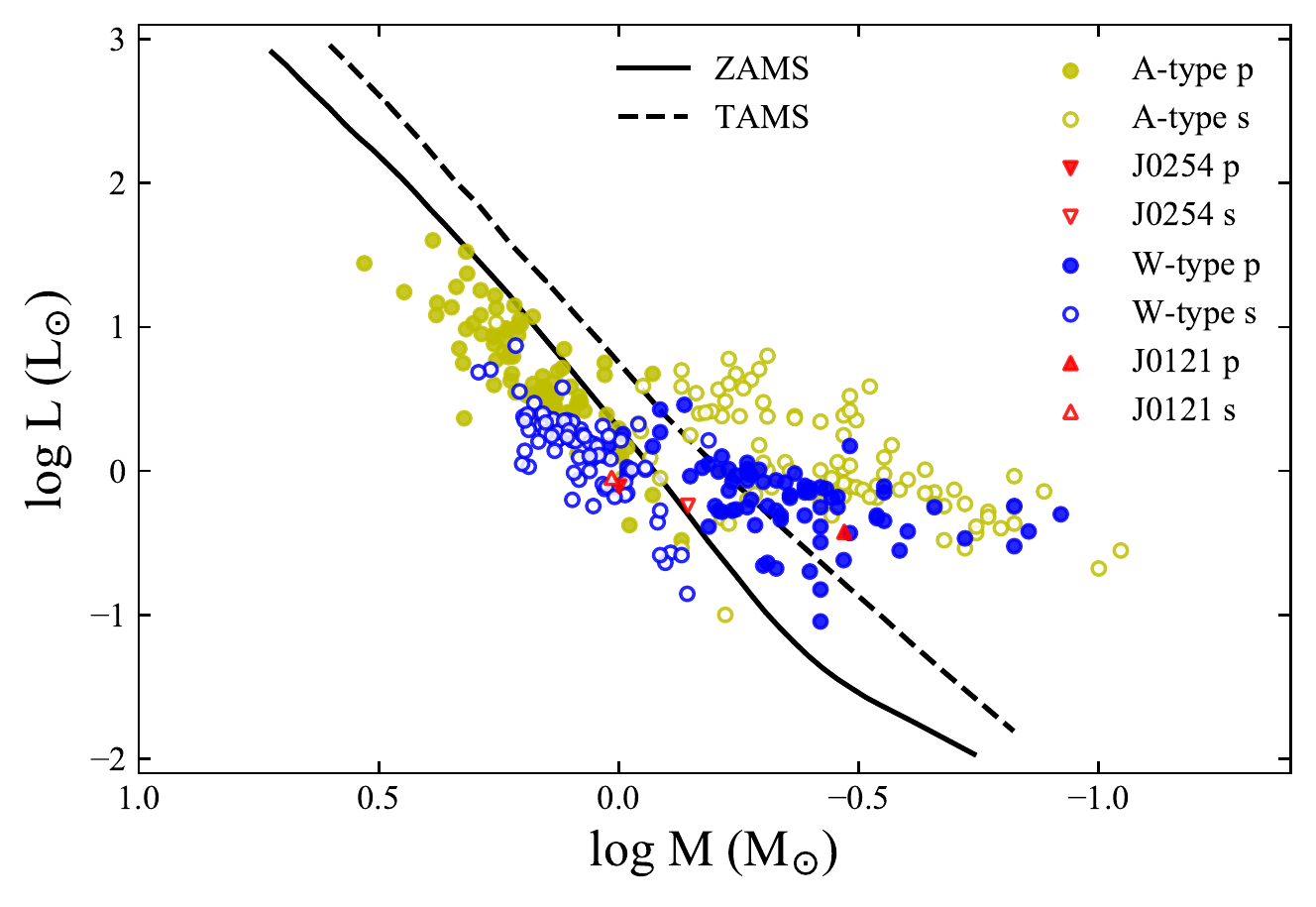}
	\caption{The mass-radius diagram (left) and mass-luminosity diagram (right). The solid line represents the zero-age main sequence (ZAMS), and the terminal-age main sequence (TAMS) is expressed as a dashed line. The two lines are constructed by the binary system evolution code provided by \citet{2002MNRAS.329..897H}. Red represents our own two targets, with $\targeta$ in regular triangles, and $\targetb$ in inverted triangles. The A-type contact binaries and the W-type contact binaries from \citet{2021AJ....162...13L} are also displayed as yellow circles and blue circles, respectively. In addition, solid triangles represent the primary components with the high temperature, which are represented by p.}
	\label{fig:mass-}
\end{figure*}

To compare with earlier studies, we used Table 7 of \citet{2021AJ....162...13L}, including 94 A-type systems and 85 W-type systems, to understand the evolutionary state of our two contact binaries. The M-R (mass-radius) and M-L (mass-luminosity) relations are plotted in Fig. \ref{fig:mass-}. As seen in the figure, in most binary systems, including both targets, the more massive part is closer to the ZAMS than the less massive part, which is different from our perception that the more massive single stars evolve faster. This phenomenon can be explained by two conjectures, one being that mass transfer leads to the mass ratio reversal \citep{1988ASIC..241..345G}, and the other is the theory of thermal relaxation oscillations, where the transfer of energy from the massive component to the less massive component increases the radius and luminosity of the less massive component, and makes its own radius and luminosity decrease. $\targeta$ is more suitable for the second conjecture explanation, while the lower-mass component of $\targetb$ has evolved faster and has left TAMS. Therefore, $\targetb$ is more suitable for the first explanation, the current lower-mass component is the initially higher-mass component of the binary system, and this component underwent a rapid mass transfer process when it left the main sequence phase.

The ratio of the spin angular momentum $J_{\text{spin}}$ to the orbital angular momentum $J_{\text {orb }}$ reflects the stability of the binary evolution. When $J_{\text{spin}} / J_{\text{orb}}<1/3$, the binary is in a relatively  stable evolutionary state. \citet{2015AJ....150...69Y} provided the equation $\frac{J_{\text {spin }}}{J_{\text {orb }}}=\frac{1+q}{q}\left[\left(k_{1} r_{1}\right)^{2}+\left(k_{2} r_{2}\right)^{2} q\right]$ which allows calculate this ratio, where $r_i$ is ratio of the radius $R_i$ to the semi-major axis of the orbit $a$, and the value of $k^2_i$ is set as 0.06 \citep{2006MNRAS.369.2001L}.According to the above equation, we obtained the ratio of the spin angular momentum $J_{\text {spin}}$ to the orbital angular momentum $J_{\text {orb }}$ for the two targets as 0.0378 ($\targeta$) and 0.0637 ($\targetb$), respectively. The results indicate that $\targeta$ and $\targetb$ are currently in a stable evolutionary state.

\begin{figure}[htbp]
	\centering
	\includegraphics[width=0.48\textwidth]{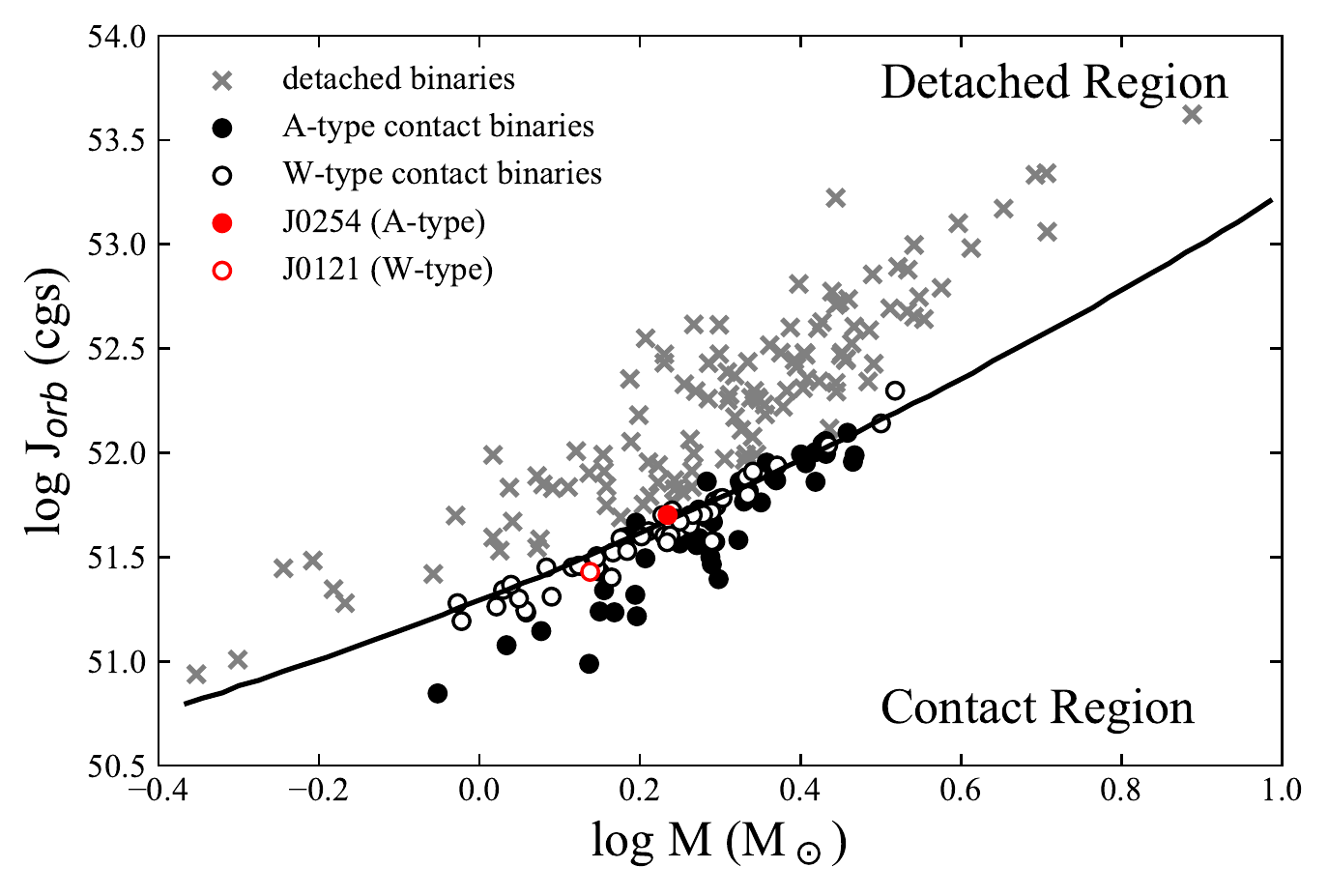}
	\caption{The mass and the orbital angular momentum diagram. The detached binaries, the A-type contact binaries and the W-type contact binaries from \citet{2006MNRAS.373.1483E} are shown as crosses in gray, black solid circles and black hollow circles, respectively. The border line in solid black, which separates the detached and contact systems, is also found by \citet{2006MNRAS.373.1483E}. The red circles represent our targets.}.
	\label{fig:logM-logJ}
\end{figure}

\citet{2006MNRAS.373.1483E} provided the correlation between the contact state and the orbital angular momentum of the binary. To investigate the contact state of the two targets, we calculate the orbital angular momentum with the equation $ J_{\text {orb }}=1.24 \times 10^{52} \times M ^{5 / 3} \times P^{1 / 3} \times q \times(1+q)^{-2} $ given by \citet{2013AJ....146..157C}, where $M$ is the total mass of the system. The orbital angular momentums of our two targets were obtained, with 51.702 ($\targeta$) and 51.429 ($\targetb$). The relationship between $\log$ $J_{\text{orb}}$ and $\log$ $M$ of two targets are shown in Fig. \ref{fig:logM-logJ}. It can be seen that $\targetb$ is a contact binary, while $\targeta$ is a shallow contact binary, which may have evolved from short-period detached binaries by angular momentum loss.

\section{Summary} \label{sec:summary}

We performed new photometric observations of $\targeta$ and $\targetb$ using NOWT. We analyzed the chromospheric activity of the two targets based on LAMOST spectroscopic data. Combining the photometric data from NOWT and the survey database, we provided the linear ephemeris of the two targets, studied their orbital period variations, and also analyzed the light curves of the two targets. We also discuss various absolute parameter estimation methods and apply them to two targets to obtain their absolute parameters. In this paper, we compare the distances of the targets with those in the literature by calculating them.

\begin{enumerate}
	\item According to the results of the light curve analysis, $\targeta$ is a typical A-type W UMa contact binary, and $\targetb$ is a W-type W UMa contact binary.
	\item The results show the presence of stellar spots with chromospheric activity in both targets, which indicates the presence of magnetic activity in $\targeta$ and $\targetb$.
	\item The $(O-C)$ analysis shows that both $\targeta$ and $\targetb$ exhibit a long-term increasing trend in their orbital periods, implying that both targets are transferring mass from their less massive companions to their more massive ones.
	\item We also studied the evolutionary status of these two targets and found that both of them are in the evolutionary stability stage. Based on the thermal relaxation oscillation theory, we predict that these two targets will transition from the contact state to the semi-detached state.
\end{enumerate}


\begin{acknowledgements}
We thank the anonymous referee very much for the very helpful comments.
This work received the generous support of the National Natural Science Foundation of China under grants U2031204. We gratefully acknowledge the science research grants from the China Manned Space Project with NO.CMSCSST-2021-A08.
We acknowledge the support of the staff of the Nanshan One-meter Wide-field Telescope (NOWT).
The spectral data were provided by Guoshoujing Telescope (the Large Sky Area Multi-Object Fiber Spectroscopic Telescope, LAMOST). LAMOST is operated and managed by the National Astronomical Observatories, Chinese Academy of Sciences.
This work includes data collected by the CRTS mission, ASAS-SN mission and ZTF mission.
This paper makes use of data from the DR1 of the WASP data (Butters et al. 2010) as provided by the WASP consortium, and computational resources supplied by the project "e-Infrastruktura CZ" (e-INFRA CZ LM2018140) supported by the Ministry of Education, Youth and Sports of the Czech Republic.
This research has made use of the International Variable Star Index (VSX) database, operated at AAVSO, Cambridge, Massachusetts, USA.
This work has made use of data from the European Space Agency (ESA) mission {\it Gaia} (\url{https://www.cosmos.esa.int/gaia}), processed by the {\it Gaia} Data Processing and Analysis Consortium (DPAC, \url{https://www.cosmos.esa.int/web/gaia/dpac/consortium}). Funding for the DPAC
has been provided by national institutions, in particular the institutions participating in the {\it Gaia} Multilateral Agreement.
This research also has made use of the SIMBAD database, operated at CDS, Strasbourg, France, NASA Astrophysics Data System Abstract Service.
Funding for the TESS mission is provided by NASA's Science Mission directorate.
This research made use of Lightkurve, a Python package for Kepler and TESS data analysis \citep{2018ascl.soft12013L}.

Software: Lightkurve \citep{2018ascl.soft12013L}, Astropy \citep{2013A&A...558A..33A, 2018AJ....156..123A}, Astroquery\citep{2019AJ....157...98G}, Tesscut \citep{2019ascl.soft05007B}, Numpy \citep{harris2020array}, Matplotlib \citep{Hunter:2007}, OCFit \citep{2019OEJV..197...71G}, PHEW \citep{munazza_alam_2016_47889}, PARSEC \citep[v1.2s][]{2012MNRAS.427..127B}, IRAF \citep{1986SPIE..627..733T,1993ASPC...52..173T}

\end{acknowledgements}

\bibliographystyle{raa}
\bibliography{bibtex}

\end{document}